\documentclass[lettersize,journal]{IEEEtran}
\usepackage{amsmath,amsfonts}
\usepackage{algorithmic}
\usepackage{graphicx}
\usepackage{textcomp}
\usepackage{xcolor}
\usepackage{multirow}
\usepackage{booktabs}
\usepackage{comment}
\usepackage{subfigure}
\usepackage{amsmath}
\usepackage{xspace}
\usepackage{balance}
\usepackage{todonotes}
\usepackage{caption} 
\usepackage[hidelinks]{hyperref}

\newcommand{\attack}{\texttt{ChargeX}\xspace}

\ifCLASSINFOpdf
\else
\fi
\begin{document}
%
\title{ChargeX: Exploring State Switching Attack on Electric Vehicle Charging Systems}
%
%
%

\author{Ce~Zhou,
        ~Qiben~Yan,
        ~Zhiyuan~Yu,
         ~Eshan~Dixit,
        ~Ning~Zhang,
        ~Huacheng~Zeng,
        ~and~Alireza~Safdari~Ghanhdari
\thanks{C. Zhou, Q. Yan (corresponding author), E. Dixit and H. Zeng are with the Department of Computer Science and Engineering, Michigan State University, East Lansing,
MI, 48823 USA e-mail: zhouce@msu.edu, qyan@msu.edu, dixitesh@msu.edu, hzeng@msu.edu.}
\thanks{Z. Yu and N. Zhang are with the Department of Computer Science and Engineering,
Washington University in St. Louis, St. Louis, MO 63130 USA e-mail: yu.zhiyuan@wustl.edu, zhang.ning@wustl.edu.}
\thanks{A.S. Ghanhdari was with the Department of Computer Science and Engineering Texas A\&M University, College Station, Tx, 77843 USA e-mail: alireza.safdari@tamu.edu.}
}

\maketitle

\begin{abstract}
Electric Vehicle (EV) has become one of the promising solutions to the ever-evolving environmental and energy crisis. The key to the wide adoption of EVs is a pervasive charging infrastructure, composed of both private/home chargers and public/commercial charging stations. The security of EV charging, however, has not been thoroughly investigated. This paper investigates the communication mechanisms between the chargers and EVs, and exposes the lack of protection on the authenticity in the SAE J1772 charging control protocol. To showcase our discoveries, we propose a new class of attacks, \attack, which aims to manipulate the charging states or charging rates of EV chargers with the goal of disrupting the charging schedules, causing a denial of service (DoS), or degrading the battery performance. \attack inserts a hardware attack circuit to strategically modify the charging control signals. We design and implement multiple attack systems, and evaluate the attacks on a public charging station and two home chargers using a simulated vehicle load in the lab environment. Extensive experiments on different types of chargers demonstrate the effectiveness and generalization of \attack. Specifically, we demonstrate that \attack can force the switching of an EV's charging state from ``stand by" to ``charging", even when the vehicle is not in the charging state. We further validate the attacks on a Tesla Model 3 vehicle to demonstrate the disruptive impacts of \attack. If deployed, \attack may significantly demolish people's trust in the EV charging infrastructure. 

\end{abstract}


%
\IEEEpeerreviewmaketitle

\section{Introduction}

Environmental pollution and energy crisis have become two of the most serious global issues, jeopardizing the health of human civilization~\cite{tie2013review}. With the transportation sector being one of the major contributors, electric vehicles (EVs) have emerged as one of the most promising solutions. Specifically, powered by the electric motors for propulsion, EVs enable sustainable transportation due to their unique advantages over conventional gas-powered automobiles for achieving zero-emission and improved energy efficiency~\cite{larminie2012electric}. In recent decades, EVs have witnessed a rapid growth and started bringing radical changes to our lives, from private commute to public transportation~\cite{editorial2020electricbus}, from freight delivery~\cite{zoe2020california} to construction industry~\cite{electric2020construction}. 
A market research company recently predicts that the EV market will reach 233.9 million units and \$2,495.4 billion by 2027, growing at a compound annual growth rate of 33.6\% in the forecast period of 2020 through 2027~\cite{meticulous2021electric}.

Recognizing the business opportunity, the EV charging infrastructure has flourished. EV chargers, a.k.a., \emph{electric vehicle supply equipment (EVSE}), can be mainly categorized into two different types: alternating current (AC) charger and direct current (DC) charger~\cite{wallbox}. The primary difference between these two charging modes lies in the location of the component where the AC power gets converted. Concretely, in the AC charging, the AC power is generally converted by an on-board charger which adds a significant weight to an EV \cite{On_off_board}, whereas the AC power is converted off-board in the DC charging system. 
Moreover, the charging rate of AC charging is substantially lower than that of DC charging due to the lower capacity of the on-board AC charger on the EV,
which can only draw limited power from the grid~\cite{wallbox}. At present, almost all of the home chargers on the market are AC chargers, while only a handful of charging stations support DC charging, most notably, Tesla Supercharger stations. 

\begin{figure}[t]
\centering
\includegraphics[width=0.45\textwidth]{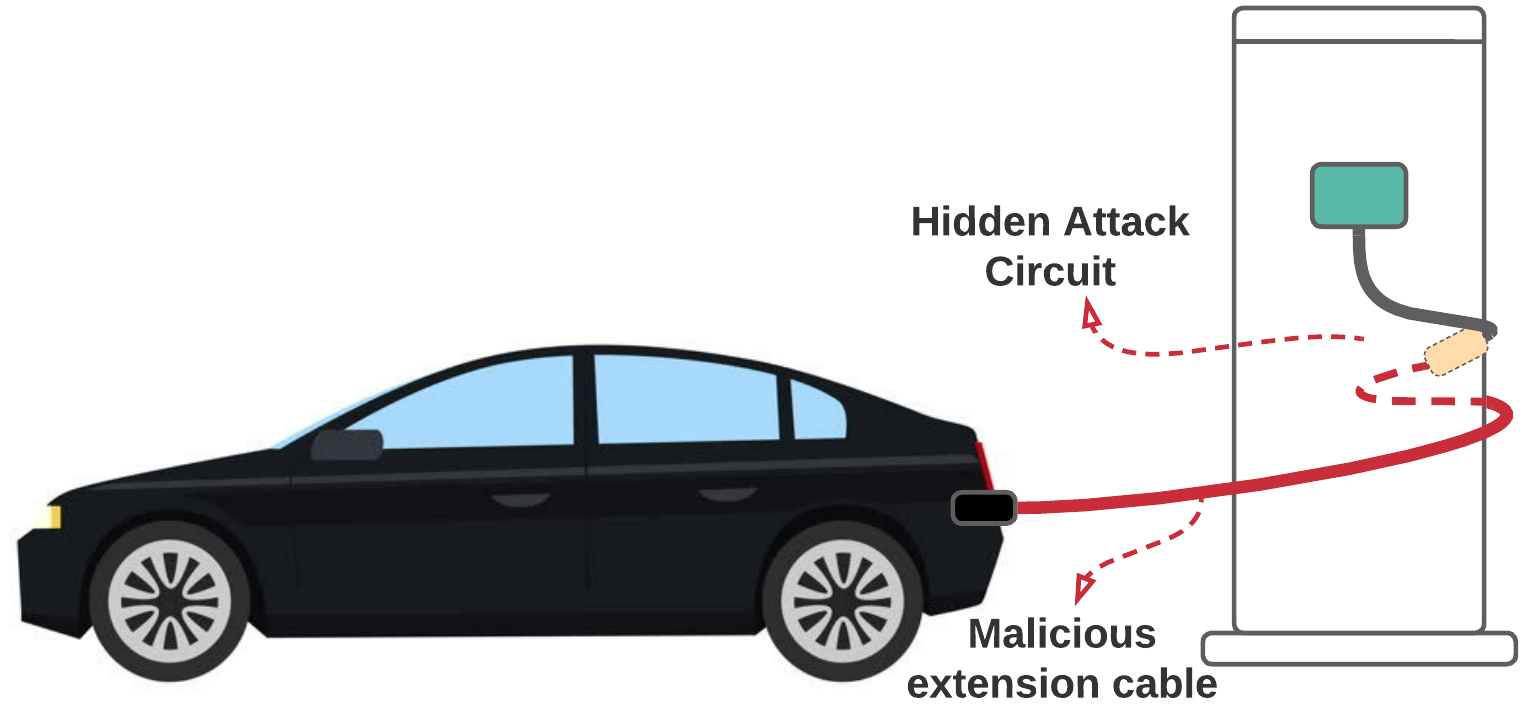}
\caption{\attack attack overview.}
\label{fig:attack_overview}
\end{figure}

The community produced a wealth of security research on EV systems over the years~\cite{yan2016can, nassi2020phantom, jin2022pla, zhou2022doublestar, xieaccess}.
Recent demonstrations of attacks on EV chargers have raised growing concerns on the resilience and robustness of modern EV systems in an adversarial environment.
For instance, a recent study~\cite{dayanikli2020electromagnetic} illustrates that the sensors and actuators in AC/DC power converters can be manipulated by intentional electromagnetic interference (IEMI) attacks.
The IEMI attacker can intentionally inject malicious electromagnetic (EM) signals into the victim's circuit to manipulate the commands sent to the actuators~\cite{selvaraj2018intentional} or to falsify the sensor readings to mislead the feedback system~\cite{tu2019trick}. Another recent study~\cite{baker2019losing} presents a passive EM-based eavesdropping attack towards DC charging that captures and interprets the digital communication between the EV and charger to recover the private messages. 
However, the existing work mainly focuses on DC charging while leaving the security investigation of AC charging \emph{an open problem}.
In fact, the AC charging systems, including the public charging stations and home chargers, have been widely deployed~\cite{chargerratio, wallbox}. The existing attacks towards DC charging cannot be directly applied to attack AC charging due to their different system designs. 
Therefore, new attack surfaces and attack approaches towards the AC charging systems need to be further explored.

In this paper, we propose \attack, a new attack against AC charging systems. Specifically, due to the widespread use of J1772 chargers in North America, we focus on identifying the vulnerabilities of the communication protocol between J1772 chargers~\cite{TexasInstruments} and EVs, and discover new attack opportunities. \attack directly modifies the communication signals by inserting a hardware attack circuit into the communication line to realize flexible state switching attacks with the potential charging rate reduction capability. 

\begin{figure}[t]
\centering
	\includegraphics[width=0.38\textwidth]{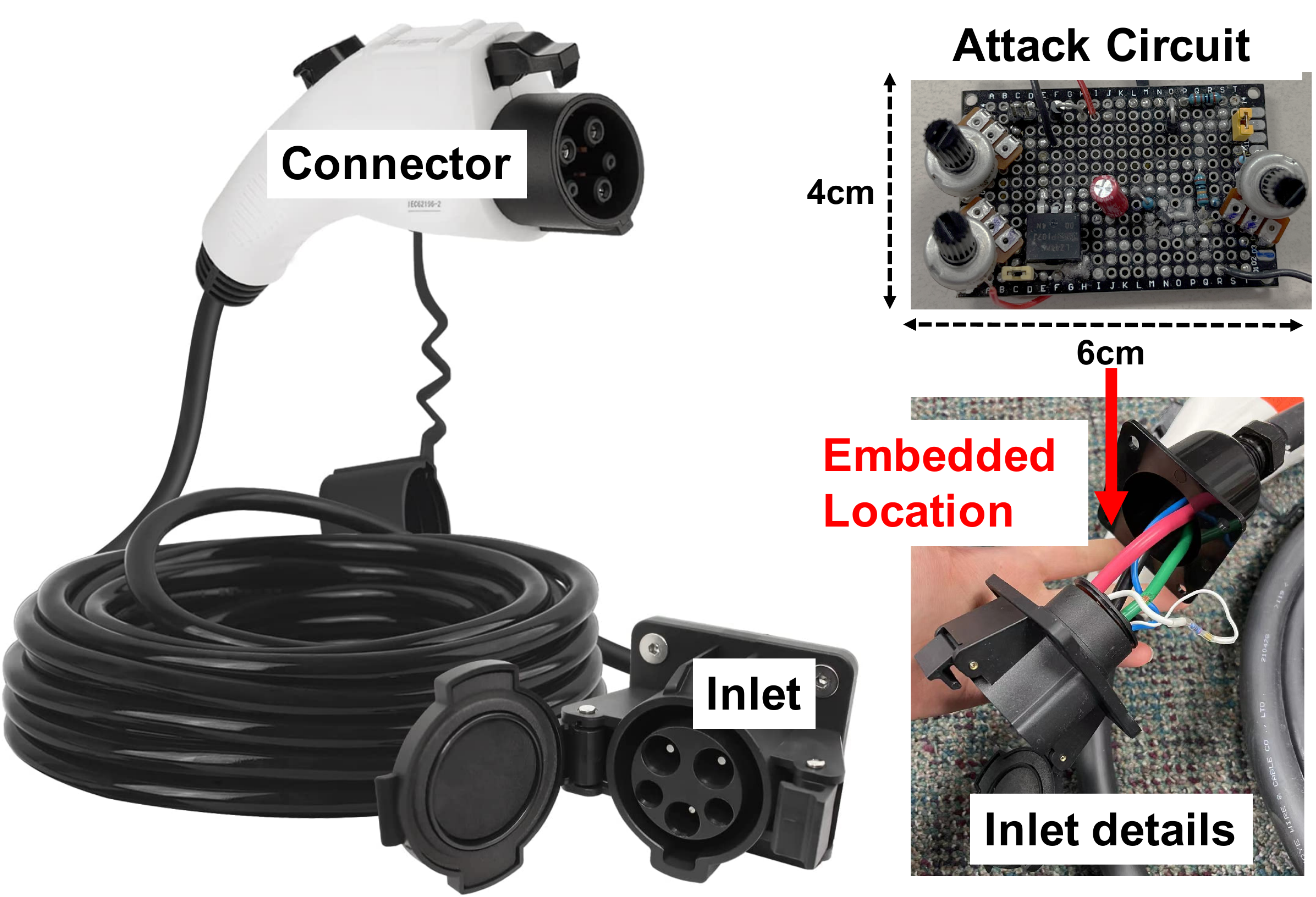}
	\caption{A J1772 extension cable with the attack circuit.
	}
	\label{fig:extension_cabl}
\end{figure}

Figure~\ref{fig:attack_overview} gives an example of \attack. The vehicle is being charged at a public charging station. Our proposed attack can prevent EV charging, overcharge EV, reduce charging rates, etc. The red cable represents the malicious J1772 extension cable \cite{extension_cable} in which the attack embeds the attack circuit. Particularly, the attack circuit is placed in the inlet of the extension cable as shown in Figure \ref{fig:extension_cabl}. The attack is covert and difficult to discover without extra caution, as the malicious extension cable can perfectly resemble that from a legitimate charger. Meanwhile, due to the small size of the attack circuit, it can be completely hidden inside the malicious cable. The connection joint can also be easily hidden, e.g., behind the charging station.

There are three major challenges in realizing \attack: (1) How to achieve the state switching? (2) How to automate the attack procedure? (3) How to manipulate the charging rate during charging? To address the first challenge, we investigate the charging control pilot signal and propose serial insertion attack and parallel attachment attack to manipulate the pilot signals which force state switching. For the second challenge, we propose an automation attack circuit design to smartly control the state transition without human involvement. The third challenge is conquered by two novel designs of the charging rate reduction circuits for adjusting the charging rate. 
 
We design hardware attack circuits in \attack, which do not require extra bulky power sources, making it suitable to be hidden in the charger cables or extension cables. 
\attack consists of state switching attack and charging rate attack, which aim at modifying the charging states and charging rates without being noticed by the EV owners. There are three hardware attack designs in state switching attack, including serial insertion attack, parallel attachment attack and automation attack. Meanwhile, two different attack designs, TLC555-based and fake load-based duty cycle attacks, belong to charging rate attack.

To evaluate the performance of \attack, we perform the attacks on one public EV charging station and two home EV chargers using a simulated vehicle load circuit in the lab environment. Regarding the real-world case study, we conduct a controlled outdoor experiment with a Tesla Model 3 2020. 
The evaluation results show that \attack can successfully switch the charging state towards the EV chargers in both lab and outdoor experimental settings. For attack demonstrations, please visit our website \textbf{\url{https://chargerattacks.github.io/}}. 

In summary, this paper makes the following contributions:

\begin{itemize}


\item We systematically analyze the communication and control mechanisms present in the J1772 EV charging protocol. We evaluate vulnerabilities at the physical layer and demonstrate the feasibility of state-switching attacks and charging rate attacks during the AC charging process.

\item We propose \attack to effectively manipulate the charging states between the EV and J1772 chargers without notifying the EV owners. Specifically, we design and implement the serial insertion attack and parallel attachment attack with human control, as well as an automation attack to realize the attacks without human involvement.

\item We identify the charging rate attack by changing the duty cycle of the control signal. We design and implement TLC555-based and Fake Load-based attacks to demonstrate the potential of the attacks for manipulating the charging rates. 


\item We evaluate the attack effectiveness in a lab environment using a simulated vehicle, as well as in real-world testing using a Tesla Model 3. We successfully demonstrate state switching attacks on both home chargers and public charging stations in real-world conditions, and carry out charging rate attacks on home chargers within the lab environment.

\end{itemize}

\section{Background} \label{background}

Understanding the working principles of EV charging is the first step towards identifying exploitable attack surfaces and developing novel attacks. In this section, we will introduce the general background of EV charging and J1772 chargers, the vulnerability assessment at the physical layer, as well as the PWM control pilot
of the EV charging standards.

\subsection{Basics of EV charging}


In the charger market, there are four major AC charging systems, including the Society of Automotive Engineers (SAE) J1772\footnote{An open standard in North America.}, Mennekes\footnote{An open standard used in Europe.}, GB/T\footnote{A nationwide standard in China.}, and Tesla charger\footnote{A proprietary standard developed by Tesla Motors.}. Meanwhile, the four most popular DC charging systems are Combined Charging System (CCS)\footnotemark[3], CHAdeMO\footnote{An open standard developed by Nissan and dominant in Japan.}, GB/T, and Tesla supercharger~\cite{hall2017emerging}. 
This work primarily focuses on the J1772 charger due to its wide deployment across North America, which has been adopted by all major EV manufacturers except Tesla. Fortunately, the Tesla car comes with a charger adapter \cite{SAEJ1772ChargingAdapter} that allows the car to use J1772 chargers. Essentially, every EV car in the North America market is compatible with the J1772 charging standard. 
Moreover, the Type 1 CCS connector used for DC charging also follows the J1772 standard~\cite{J1772_wiki, CCS_design}.
Table \ref{tab:charging_level} displays five charging levels in the latest J1772 standard \cite{J1772_201710}: AC Level 1--3 and DC Level 1--2. 
In this paper, we mainly focus on AC Level 1 and AC Level 2 chargers due to their popularity for EV charging in the market. The study could also be applicable to the DC chargers that follow the J1772 standard.

\begin{table}[htp]
\centering
\caption{Different charging levels in J1772 AC and DC charging (North America)}
\label{tab:charging_level}
\resizebox{0.48\textwidth}{!}{%
\begin{tabular}{|l|l|l|l|}
\hline
Levels & \begin{tabular}[c]{@{}l@{}}AC - Nominal Supply ($V$)\\ DC - EVSE Output ($V$)\end{tabular} & \begin{tabular}[c]{@{}l@{}}Max Current \\ ($A$) \end{tabular}  & \begin{tabular}[c]{@{}l@{}}Max Power \\ ($kW$) \end{tabular} \\ \hline
AC Level 1 & $120$       & $12$ or $16$ & $1.44$ or $1.92$ \\ \hline
AC Level 2 & $208 - 240$ & $24 - 80$    & $5.0 - 19.2$     \\ \hline
AC Level 3 & $208 - 600$ & $63 - 160$   & $22.7 - 166$     \\ \hline
DC Level 1 & $50 - 1,000$ & $80$         & $80$             \\ \hline
DC Level 2 & $50 - 1,000$ & $400$        & $400$            \\ \hline
\end{tabular}%
}
\end{table}

\begin{figure}[t]
\centering
 	\includegraphics[width=0.5\textwidth]{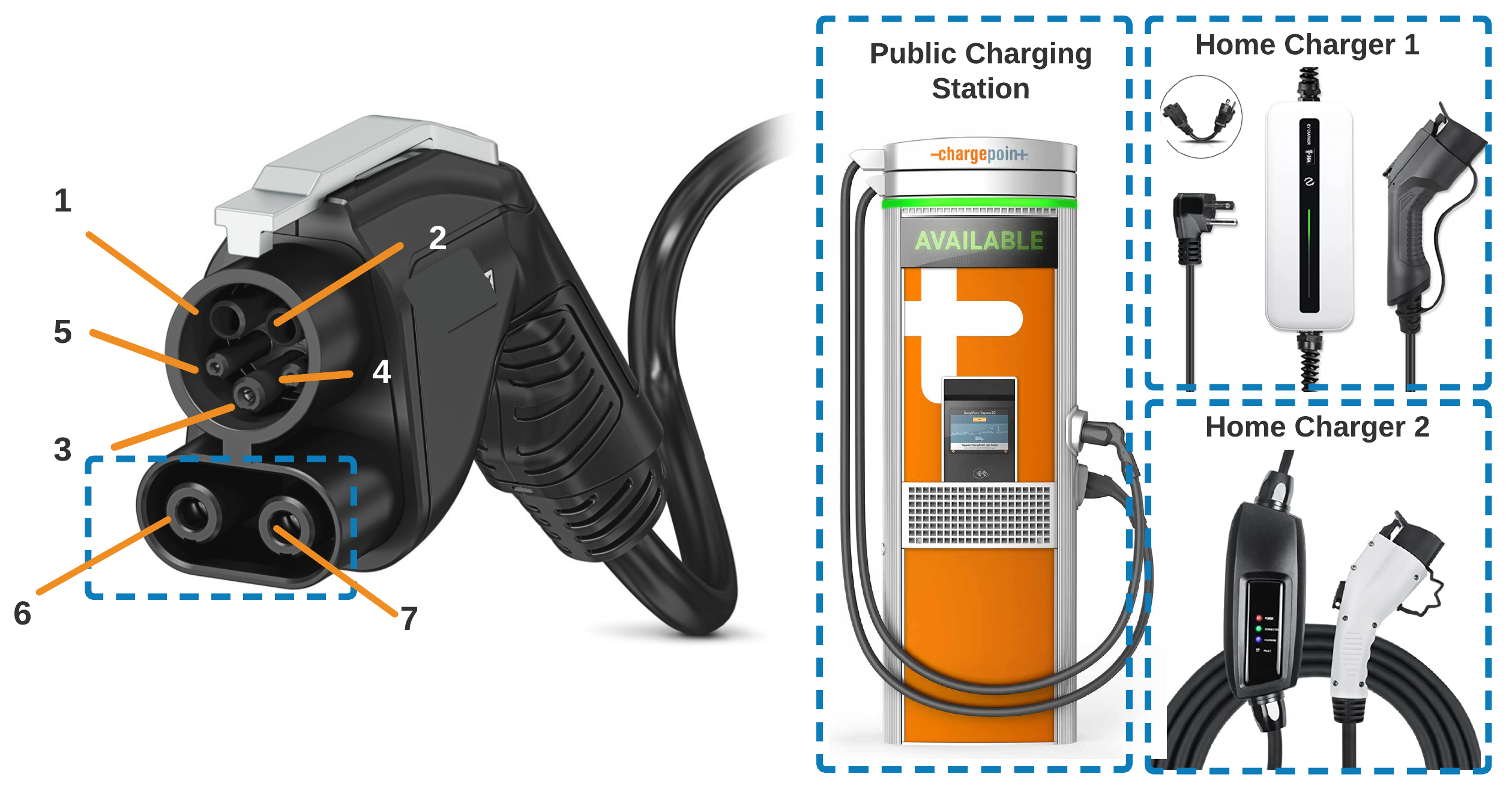}
	\caption{The DC combination connector (CCS Combo1) and the chargers. With the removal of DC Pins 6 and 7, CCS Combo1 becomes a Type 1 ``J1772" (Japan/US) AC connector. The three chargers include a public charging station and two home chargers. 
	}
	\label{fig:connector}
\end{figure}

\subsection{Vulnerability Assessment at the Physical Layer}\label{Communication_Protocol}

In order to assess the security of EV charging, it is necessary to study the communication protocols of EV charging within the context of potential cyber attack scenarios. Figure~\ref{fig:connector} and Table~\ref{tab:pin_function} show the pin configuration of the standard J1772 connector. 
Pins 1 and 2, \emph{i.e.}, L1 and L2, supply the DC power for Level 1 and Level 2 chargers. Equipment Ground (GND) provides ground to the power source, Proximity Detection (PD) verifies whether or not a charger is connected to the vehicle, and Control Pilot (CP) is the primary control conductor, with which the EVSE carries out two different types of communication with the EV. The first type is a \emph{pulse-width modulated (PWM)}
\emph{control pilot}, through which the vehicle's requested charging current information can be delivered to the charger. This is a low-level communication that is IEC 61851 standardized.
The other type is \emph{digital data transfer},  which provides powerline
communication (PLC) for exchanging information in DC charging mostly, including state of charging, remaining charging time, payment information, etc. This high-level communication is standardized by ISO 15118. Digital data transfer is optional in AC charging. Note that the PLC shares the lines with the IEC 61851 signaling system with the signals superposed at the physical layer. These two types of communication are the primary attack targets in EV charging attacks at the physical layer. 
Given that both AC and DC charging involve low-level PWM control pilot communication, the exploitation of PWM control pilot potentially opens up avenues for attacks on both AC and DC charging processes.
The two pins at the bottom of Figure~\ref{fig:connector}, \emph{i.e.}, Pins 6 and 7,  provide the DC power for DC level 2 charging. 
Without these two pins, the pin layout becomes that of the connector of AC Level 1 and Level 2 chargers.

\begin{table}[htp]
\centering
\caption{Pin functions in AC and DC charging}
\label{tab:pin_function}
\resizebox{0.37\textwidth}{!}{%
\begin{tabular}{|l|ll|}
\hline
\multirow{2}{*}{Pin No.} & \multicolumn{2}{l|}{Connector Function}                                                                                     \\ \cline{2-3} 
  & \multicolumn{1}{l|}{AC} & DC                   \\ \hline
1 & \multicolumn{1}{l|}{L1} & DC Level 1 Power (+) \\ \hline
2                        & \multicolumn{1}{l|}{\begin{tabular}[c]{@{}l@{}}Neutral - AC Level 1 \\ L2 - AC Level 2\end{tabular}} & DC Level 1 Power (-) \\ \hline
3 & \multicolumn{2}{l|}{Equipment Ground (GND)}    \\ \hline
4 & \multicolumn{2}{l|}{Control Pilot (CP)}        \\ \hline
5 & \multicolumn{2}{l|}{Proximity Detection (PD)}  \\ \hline
6 & \multicolumn{1}{l|}{-}  & DC Level 2 Power (+) \\ \hline
7 & \multicolumn{1}{l|}{-}  & DC Level 2 Power (-) \\ \hline
\end{tabular}%
}
\end{table}

\subsection{PWM Control Pilot}\label{PWM_Control_Pilot}

\begin{figure}[htp]
\centering
	\includegraphics[width=0.5\textwidth]{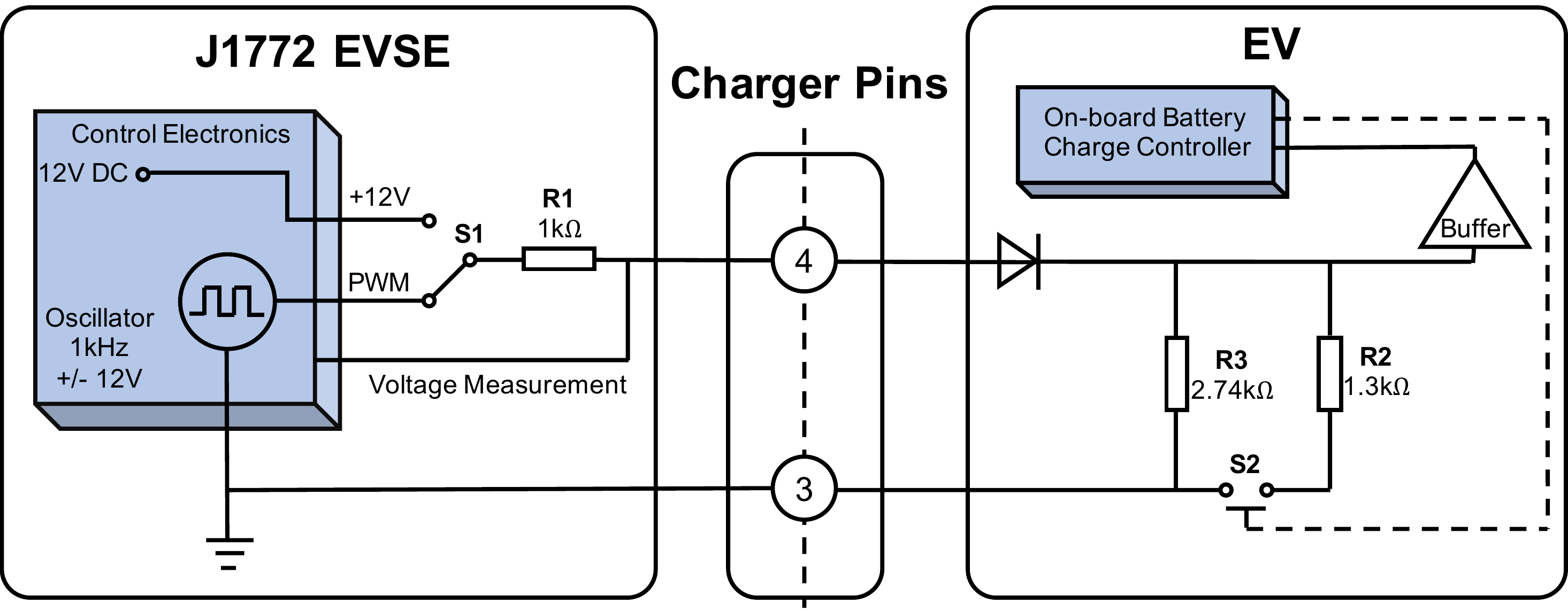}
	\caption{A typical J1772 EVSE control circuit.}
	\label{fig:typical_cp_circuit}
\end{figure}

Our work focuses on the state switching in EV charging procedure, which is reflected in the PWM control pilot signals. 
Specifically, Figure~\ref{fig:typical_cp_circuit} shows a typical EVSE control circuit for state switching, and Table~\ref{tab:EVstate} lists the different charging states. 
States A, B, and C are the core charging states,
which represent the normal charging operations from EV detection to EV charging. 
The occurrence of various types of charging errors is reflected in states E and F.

In Figure \ref{fig:typical_cp_circuit}, $R1$ is the resistor in the charger interface that creates a voltage divider network 
 for pilot signal measurement, 
whereas $R2$ and $R3$ are the load resistors used for state switching on the EV side. The standard EV charging procedure consists of the following steps: (1)  EVSE supplies $12V$ DC to the pilot line, corresponding to the state A. As soon as the charging plug is connected to EV, the pilot DC notifies EV of the connection. (2) Once the plug is properly connected, $S1$ is immediately switched to output a $\pm 12V$ PWM signal,  
while the EV puts a $2.74k\Omega$ load (\emph{i.e.}, $R3$) to the pilot line. The load resistor will lower the high voltage level of pilot signals to $9 V$. 
(3) EVSE detects the PWM voltage drop and enters state B, 
which subsequently informs the vehicle about the amount of current it can draw. The EVSE is now ready to supply power to the vehicle. (4) The EV switches to the $822\Omega$ load by closing $S2$ switch (with $R2$ connected) and starts drawing the power. 
This will reduce the PWM high voltage level to $6V$ and notify the EVSE that the charging process has begun (state C). (5) 
The charging process is terminated when the cable is unplugged or an error is detected, in which case the pilot line voltage returns to $12V$ DC. 
The EVSE then terminates the charging and returns back to state A or falls into an error state (states E or F).

\begin{table}[htp]
\centering
\caption{J1772 EVSE charging states}
\label{tab:EVstate}
\resizebox{0.49\textwidth}{!}{%
\begin{tabular}{|l|l|l|l|l|l|}
\hline
State &
  \begin{tabular}[c]{@{}l@{}} High Pilot\\ Voltage (V)\end{tabular} &
  \begin{tabular}[c]{@{}l@{}}Low Pilot  \\ Voltage (V)\end{tabular} &
  \begin{tabular}[c]{@{}l@{}}Pilot \\ Freq. (kHz)\end{tabular} &
  Resist. ($\Omega$) &
  EV Status \\ \hline
State A & $12$ & N/A    & DC     & N/A       & Not connected \\ \hline
State B & $9$  & $-12$ & $1$ & $2,740$ & EV detected \\ \hline
State C & $6$  & $-12$ & $1$ & $882$   & EV charging \\ \hline
State D & $3$  & $-12$ & $1$ & $246$   &  With ventilation \\ \hline
State E & $0$  & $0$   & N/A    & —         & No power \\ \hline
State F & N/A    & $-12$ & N/A    & —         & Error \\ \hline
\end{tabular}%
}
\end{table}

The EVSE delivers the maximal available current capacity to the EV by adjusting the pilot duty cycle, which is the percentage of a period when the signal is high. The EV then uses the duty cycle of PWM to control the AC current of the on-board charger, which is
drawn from the power line. 
The relationship between the supply current and the duty cycle of PWM signals is presented in Table~\ref{duty_cycle_table}.
Specifically,  a duty cycle of $3$-$7$\% will be interpreted as a command for digital communication. 
If the duty cycle is $\leq$85\%, the current reading follows the  formula: $Amp=(\%dutycycle)\cdot 0.6$. If the duty cycle is $>$85\%, 
the maximal current reading becomes 
$Amp=(\%dutycycle-64)\cdot 2.5$. If the duty cycle exceeds 96\%, the EV considers it as a valid 96\% duty cycle, which corresponds to the maximum current $80A$. 

\begin{table}[htp]
\centering
\caption{The relationship of the supply current and the duty cycle of PWM signals}
\label{duty_cycle_table}
\resizebox{0.43\textwidth}{!}{%
\begin{tabular}{|l|l|l|l|}
\hline
Duty Cycle (\%) & Current (A) & Duty Cycle (\%) & Current (A) \\ \hline
10   & 6    & 25   & 15 \\ \hline
16.6 & 10   & 26.7 & 16 \\ \hline
17.4 & 10.4 & 50   & 30 \\ \hline
18.4 & 11   & 53.3 & 32 \\ \hline
\end{tabular}%
}
\end{table}




\section{Threat Model}\label{subsec:threatmodel}
In this section, we present the attack goals, the attacker's capabilities, and attack scenarios of \attack.



\textbf{Attack Goals.} 
The attack goal is to manipulate the charging states to cause the undesired battery charging behavior on EV chargers, which influences the normal battery charging process.
The attackers may reduce or completely halt the battery charging by launching a DoS attack. They may also attempt to switch the charging state to overcharge and damage the battery or manipulate the charging current to significantly alter the charging rate. There are also security implications beyond the charging activity of individual EVs. An attacker who compromises multiple EV chargers may cause harmful disturbance to power delivery infrastructure or commit financial fraud by deliberately slowing down the charging speed while charging the same price. 

\textbf{Attacker's Capabilities.}
We assume the attacker can get close to the target charger, but the attacker does not have the opportunity to modify the chargers in a stealthy manner. Instead, the attacker can install a pre-assembled malicious charging extension cable as shown in Figure \ref{fig:extension_cabl}, which resembles the legitimate charging cable. 
We assume the EV owners leave the vehicle during charging so that they will not observe the alerts from the vehicles. We assume the victim charger follows the J1772 charging standard. 

\textbf{Attack Scenarios.}
Our attack can work on both public charging stations and home chargers. The attacker connects a malicious extension cable, which contains the charging state manipulation circuit, to a charging station as shown in Figure~\ref{fig:attack_overview}. Given that these attacks can be executed on public charging stations, a significant number of vehicles could potentially be affected.
When targeting a home charger, the attacker may directly replace the home charging extension cable with the malicious one. We assume that the home charger is not in a physically secured environment and it is publicly accessible. Note that the attack circuit is very tiny, which requires no additional power supply in the design.

Compared with the physical damage to the public chargers, such as hitting the stop button, unplugging the power line plugs, or cutting the transmission/power cord, our attacks are more feasible and practical. Public chargers are generally not equipped with the “stop button” or power line plugs. Even if the power line of the chargers is exposed, cutting the transmission/power cord is quite dangerous and highly noticeable due to its high power. On the other hand, fault data injection related attacks towards the high-level communication line usually require an attack circuit with a microcontroller and power supply, which needs a large footprint and maintenance. It would be difficult to fit a large attack circuit into a malicious extension cable and use it for a prolonged time period.

\section{State Switching Attack Design}\label{Sec:Attack_Design}
In this section, we present state switching attack design. 
Our attack aims to control the switching between the charging states by tampering with the charging cable hardware. 
According to Table~\ref{tab:EVstate}, since the charging states of the charger are directly determined by the high pilot voltage at the EV charging inlet, the key idea of the attack is to manipulate the high pilot voltage. 

To conceal the attack components, we use an extension cable as shown in Figure~\ref{fig:extension_cabl} to implement state switching attack. Such an extension cable is originally designed to extend the charging cable length. 
Here, we introduce three types of attack designs with different hardware modifications.

\subsection{Serial Insertion Attack}
\begin{figure}[t]
\centering
\subfigure[Serial insertion attack design]{\includegraphics[width=0.49\textwidth]{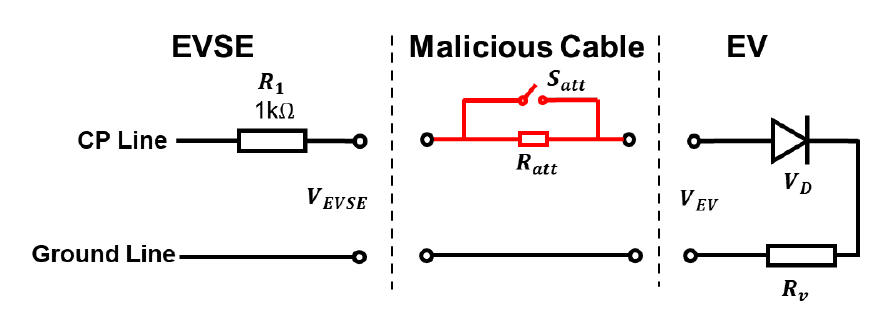}\label{fig:serial_design}}
\subfigure[Parallel attachment attack design]{\includegraphics[width=0.49\textwidth]{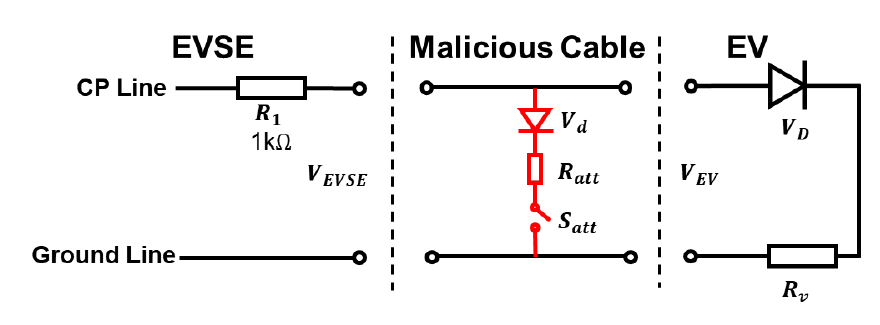}\label{fig:parallel_design}}
\caption{The design of serial insertion and parallel attachment attacks.}
\label{fig:attack_circuit_design}
\end{figure}

In the serial insertion attack, the attacker simply adds 
an additional resistor $R_{att}$ in serial with the resistor and diode inside the EV as shown in Figure~\ref{fig:serial_design}.
The attacker also deploys a switch component $S_{att}$ to control the attack operation. In general, the serial insertion attack switches the charging state of the charger to the previous states, e.g., state C to A or B, while switching the charging state  of the EV to the latter state, e.g., state B to C or F. The goal of the serial insertion attack is to introduce a different perception of the charging states at the EV and EVSE, which will lead to a communication fault, \emph{i.e.}, a DoS outcome. 

Without attaching the malicious extension cable,
$V_{EVSE}$ is equal to $V_{EV}$, \emph{i.e.}, both the EVSE and EV have the same perception of the charging state. However, with the addition of $R_{att}$, the difference between the $V_{EVSE}$ and $V_{EV}$ grows. By carefully manipulating the $R_{att}$, we can realize the state switching attack. However, if the difference grows beyond a threshold, there will be a communication error between the EV and EVSE, and as a result, the vehicle's charging state may fall into the failure state F. 

It is worth noting that each charging state has an \emph{error tolerance range}, which indicates a range of permissible errors used to withstand the environmental noise and disturbance. By carefully adjusting the $R_{att}$, a proper difference between  $V_{EVSE}$ and $V_{EV}$ allows the state switching to happen without trapping into the failure state. 
Specifically, $V_{EVSE}$ and $V_{EV}$ can be written as follows: 
\begin{equation}\label{equ1}
    V_{EVSE}=12V-1k\Omega\cdot\frac{12V}{1k\Omega+R_{att}+R_{v}},
\end{equation}

\begin{equation}\label{equ2}
    V_{EV}=12V-(1k\Omega+R_{att})\cdot\frac{12V}{1k\Omega+R_{att}+R_{v}},
\end{equation}
where $R_{v}$ represents the resistance of the resistor on the EV at a normal charging state. 
We omit the forward voltage of $V_{D}$ (Figure~\ref{fig:serial_design}) as it is usually a small value. 
In essence, when the following condition satisfies, different states are perceived by the EV and charger:
\begin{equation}\label{equ3}
    V_{Diff}=V_{EVSE}-V_{EV}\geq \lambda,
\end{equation}
where the $\lambda$ denotes the error tolerance range of normal states A, B, or C.

Let us consider state B as the benign charging state. Assume the corresponding voltage of the state switching boundary from A to B and B to C are $V_{A/B}$ and $V_{B/C}$, respectively. In order to achieve a successful attack, the difference between $V_{EVSE}$ and $V_{EV}$ should be large enough, such that the EV and the charger have different perceptions of the charging state. In other words, as long as $V_{Diff}\geq V_{A/B}-V_{B/C}$, two sides (\emph{i.e.}, EV and charger) are guaranteed to have different state perceptions, and as a result, the communication fault will occur. 


\subsection{Parallel Attachment Attack}
Similar to the serial insertion design, in order to manipulate the load resistance, 
another way to realize state switching is to attach an additional resistor in parallel with the load resistors in EV. Notably, \emph{parallel attachment attack} can switch the state of both EV and EVSE to the latter state, for instance, state B to C or F.

Particularly, the parallel resistor can reduce the total resistance of the vehicle load, which results in state switching as shown in 
Table~\ref{tab:EVstate}. As the EV uses a diode to filter out the negative voltage of the PWM signals, the added parallel circuit also includes a diode to reject the negative voltage in order to allow for a normal charging operation. The forward voltages $V_d$ and $V_D$ are negligible, and thus are neglected. 


The resistance $R_{att}$ is used to control the high voltage level of the PWM signal to achieve the state switching. 
The total resistance of the vehicle load can be written as: 
\begin{equation}
    R_{load}=\frac{1}{\frac{1}{R_{v}}+\frac{1}{R_{att}}}.
    \label{equ4}
\end{equation}
Since $V_{EVSE}$ is always equal to $V_{EV}$ in this design, we can then calculate the high voltage level of PWM signal for both sides as:
\begin{equation}
    V_{EVSE}=V_{EV}=12V-1k\Omega\cdot\frac{12V}{1k\Omega+R_{load}}.
    \label{equ5}
\end{equation}
To achieve state switching, $V_{EVSE}$ should be increased by the added parallel resistor to
cross the state switching boundary. 
For example, if an attacker aims to switch the state from state B to C, $R_{att}$ should be properly configured such that $V_{EVSE}\leq V_{B/C}$. However, if $R_{att}$ is too small, 
the charging state may fall into state F. 
Therefore, the range of $R_{att}$ should be carefully tuned, such that the following inequality satisfies: 
\begin{equation}
    V_{C/F} \leq V_{EVSE}\leq V_{B/C},
    \label{equ6}
\end{equation}
where $V_{C/F}$ denotes the corresponding voltage of the state switching boundary from C to F. 

\subsection{Attack Automation}

By adding fairly simple electrical components, the design of serial insertion and parallel attachment attacks is easily realizable. 
However, such a design requires careful profiling of the victim in order to learn about the state switching boundaries. Moreover, a human attacker must be involved in the process to turn on/off the switch to initiate, terminate, and tune the attack. 
 
To address the above limitations, we propose an automation attack design that can automatically and precisely manipulate the voltage of the pilot signals, thereby enabling a more generalized and easily-deployable attack.
Basically, the attack circuit is designed to activate the attack only when the vehicle is in  the target state. In other words, when the vehicle is not connected to the charger, the charger will display a normal state A with a pilot voltage of $12V$ DC. When the vehicle is connected, the attack circuit will automatically switch the normal state into the next one, e.g., from state B to C.

\begin{figure}[]
\centering
\includegraphics[width=0.4\textwidth]{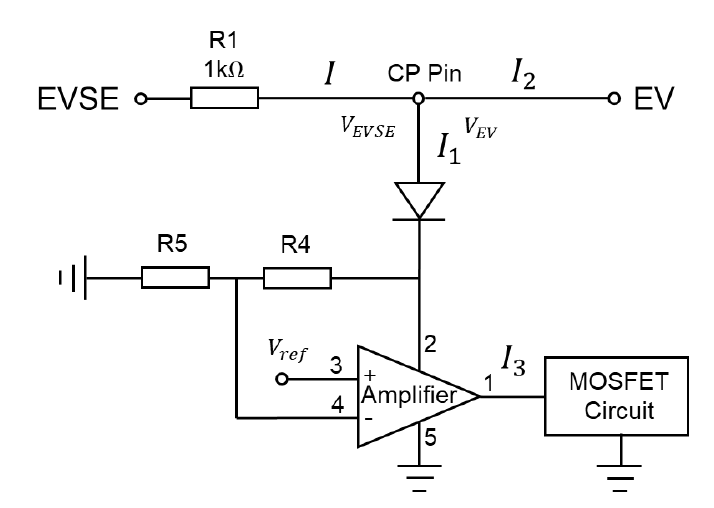}
\caption{The design of the automation circuit.}
\label{fig:auto_design}
\end{figure}

Figure \ref{fig:auto_design} shows the high-level design of the automation attack circuit. Similar to the parallel attachment attack circuit, the automation circuit is attached 
in parallel with the vehicle, such that $V_{EVSE}=V_{EV}$. 
If an attacker wishes to switch from state B to state C, the automation circuit will be activated when in state B. Without the attack, $V_{EVSE}$ is $9V$. However, since the attack circuit draws current from the pilot line, more current will flow through the $R1$ resistor. This implies that the voltage drop on $R1$ increases, subsequently leading to the state switching due to the drop of $V_{EVSE}$. In our design, we can control the voltage drop on $R1$ to be $3V$, so that $V_{EVSE}=6V$, allowing the state switching (\emph{i.e.}, B to C) to happen.  

Here, we describe the detailed design of the automation circuit, including a voltage divider, an amplifier, and a MOSFET circuit block. The amplifier and the MOSFET circuit control the activation of the automation attack. Specifically, increasing the output voltage of the amplifier will 
turn on the MOSFET circuit. Since the MOSFET circuit draws the majority of the current, 
the current flow through $R1$ will also increase, which activates the attack. 

$R4$ and $R5$ compose the voltage divider, where the voltage in-between applies to Pin 4 (negative input) of the amplifier. 
The amplifier works as a comparator that compares the voltage between the positive input (Pin 3) and the negative one. Pin 3 is connected with a voltage reference $V_{ref}$, which is a constant value (e.g., $1.2V$).  The output of the amplifier 
will be $0V$
if the voltage of the negative input is higher. Otherwise, if the positive input is higher, the output will be the same voltage as Pin 2. 

Regarding the MOSFET circuit, if there is a positive voltage (e.g., $2V$) on it, the circuit will be turned on. 
Conversely, when there is no voltage at Pin 1, the MOSFET circuit becomes an open circuit, indicating that the current $I_3\approx0A$. 
We have $I=I_1+I_2$ and $I_1\approx I_3$, as the current flowing through other branches is negligible. 
Therefore, if the MOSFET circuit is 
on, $I_1$ will increase, which causes the extra voltage drop on $R1$. 
It is worth noting that 
the voltage at Pin 4 is different in different charging states. For example, if the voltage at Pin 4 is larger than
$V_{ref}$ at state A, the attack will not be initiated and $V_{EVSE}=V_{EV}=12V$.
To launch the attack in state B, we can let the voltage at Pin 4 smaller than $V_{ref}$ by adjusting the voltage divider, $R4$ and $R5$, 
to activate the attack. As a result, $V_{EVSE}=V_{EV}=9V-3V=6V$, leading to the charging state C.

\section{State Switching Attack Evaluation}
In this section, we test the performance of the state switching attack on commercial EV chargers in the lab environment, and then we experiment the attack with chargers connected to an EV (\emph{i.e.}, Tesla).

\subsection{Experimental Setup}
\begin{figure}[htp]
\centering
\includegraphics[width=0.49\textwidth]{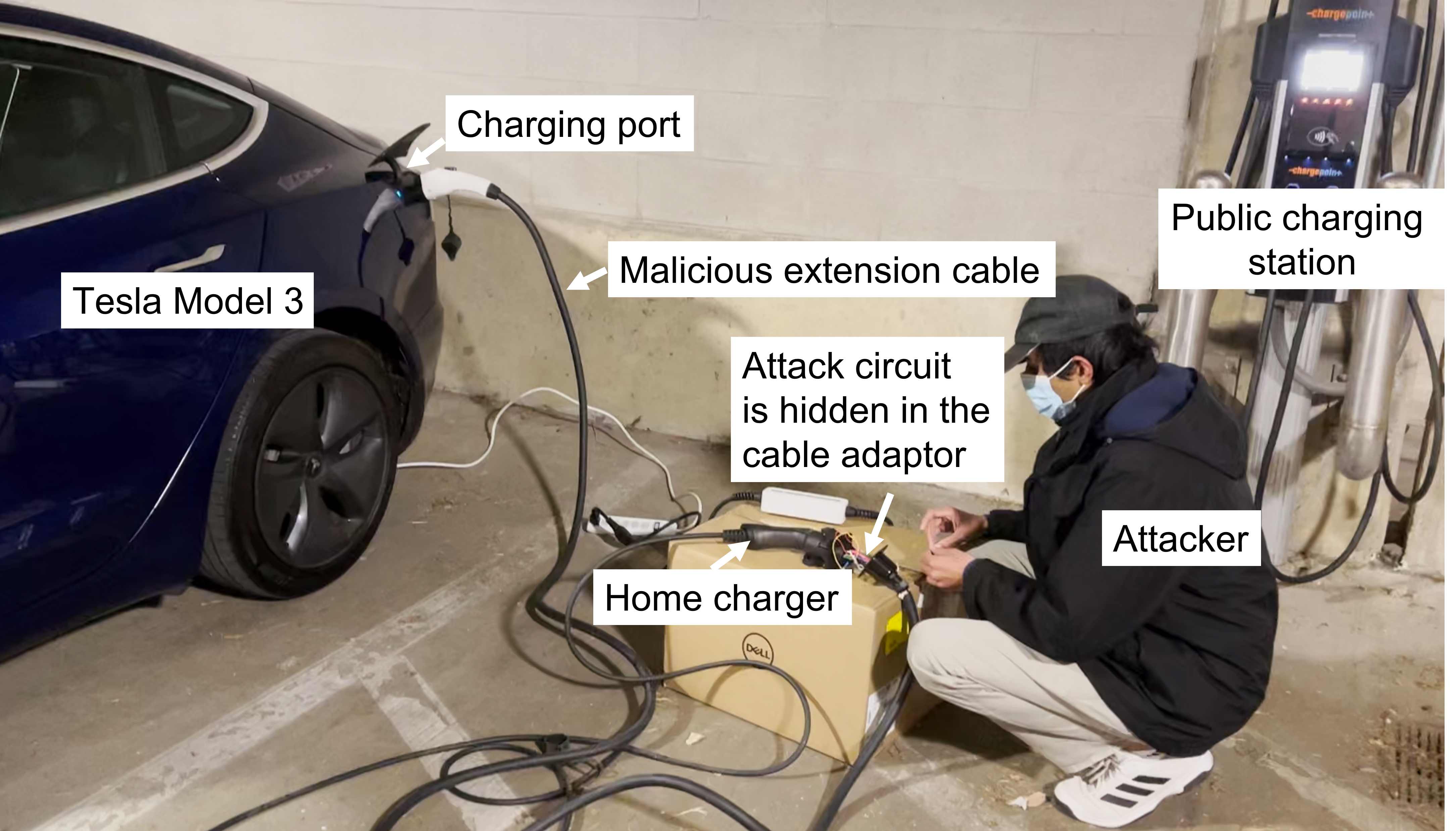}
\caption{The attack experimental setup for attacking the EV chargers 
with Tesla Model 3.}
\label{fig:Cable_attack_setup}
\end{figure}

The experimental setup is shown in Figure~\ref{fig:Cable_attack_setup}. An EV charger is connected to the malicious extension cable before being connected to the simulated vehicle load (for lab experiment) or the EV (for Tesla experiment). We 
perform the experiments on three commercial EV chargers, including AC level 1\&2 home chargers ~\cite{level1_charger, level2_charger}, and a AC level 2 public charging station (\emph{ChargePoint}~\cite{chargepoint}), denoted as Charger 1 and 2, and Public Charger hereinafter. 
The experiments are carefully controlled by only adjusting the low-power control pilot to avoid damages to either the EV or chargers. 
Charger 1 supplies $10A/16A$ and $110V/220V$, Charger 2 supplies $16A$ and $110V/220V$ for charging, 
while the Public Charger supplies $30A/220V$. 
We hide the attack circuit in the extension cable. 
The Tesla has a dedicated charging connector for both AC and DC charging. Therefore, we use a charger adaptor~\cite{SAEJ1772ChargingAdapter}, which comes standard with purchase of any Tesla vehicle. 

We test all the state switching cases in the experiments in a lab environment, while several test cases 
are omitted in the Tesla experiment due to the practicality of the attack scenarios. For example, switching any other states to state A (\emph{i.e.}, vehicle unconnected) would unlikely result in any harmful outcomes. 
To collect the attack results, in the lab test, we can observe the charging status of the charger control panel; in the real vehicle test,
the attack results are presented on both the Tesla control screen and charger control panel. 
We repeat the experiments at least three times for each case and record the average value of attack parameters for each attack. 

For the automation attack, we design our own PCB board with the size $4cm \times 6cm$, as shown in Figure~\ref{fig:extension_cabl}. The PCB can be fit into the inlet of the extension cable. Note that the PCB size can be made much smaller, if the PCB design is further optimized. 
We evaluate the attack performance by observing state switching outcomes. 

In the normal charging process, once the charger's connector is plugged into the vehicle inlet, the vehicle and the charger communicate with each other following the charging protocol (see Section \ref{PWM_Control_Pilot}). 
Therefore, the state will automatically switch from A to B, and then C. To make it stay in state B, we can press the ``stop charging" button on Tesla control panel In our experiments. 
\begin{table}[htp]
\centering
\caption{High pilot voltage thresholds for state switching boundaries}
\label{tab:state_switch_boundary}
\resizebox{0.32\textwidth}{!}{%
\begin{tabular}{|l|l|l|l|}
\hline
States Boundary                                                                                             & A/B    & B/C   & C/F        \\ \hline
\begin{tabular}[c]{@{}l@{}}Corresponding voltage \\ for Charger 1 ($V$)\end{tabular} & $10.6$ & $7.8$ & N/A \\ \hline
\begin{tabular}[c]{@{}l@{}}Corresponding voltage\\ for Charger 2 ($V$)\end{tabular}  & $10.6$ & $7.8$ & $4.4$      \\ \hline
\end{tabular}%
}
\end{table}

\begin{table*}[t]
\scriptsize
\centering
\caption{Estimated and measured $R_{attack}$ in serial and parallel design ($\leftarrow or \rightarrow$ stands for state switching direction for EVSE and EV, respectively, X$\leftarrow$ X means that the state does not switch)}
\label{tab:serial_parallel_result}
\resizebox{0.9\textwidth}{!}{%
\begin{tabular}{|l|ll|lll|}
\hline
\multirow{2}{*}{$R_{att}$} &
  \multicolumn{2}{c|}{Serial Insertion Attack} &
  \multicolumn{3}{c|}{Parallel Attachment Attack} \\ \cline{2-6} 
 &
  \multicolumn{1}{l|}{A $\leftarrow$ B $\rightarrow$ C} &
  B $\leftarrow$ C $\rightarrow$ F &
  \multicolumn{1}{l|}{B $\rightarrow$ C} &
  \multicolumn{1}{l|}{C $\rightarrow$ F} &
  B $\rightarrow$ F \\ \hline
\begin{tabular}[c]{@{}l@{}}Estimated $R_{att}$ in \\ Charger 1 ($k\Omega$)\end{tabular} &
  \multicolumn{1}{l|}{$\geq 1.0$} &
  $\geq 0.63$ &
  \multicolumn{1}{l|}{$1.68 - 5.76$} &
  \multicolumn{1}{l|}{N/A} &
  N/A \\ \hline
\begin{tabular}[c]{@{}l@{}}Estimated $R_{att}$ in \\ Charger 2 ($k\Omega$)\end{tabular} &
  \multicolumn{1}{l|}{$\geq 0.93$} &
  $\geq 0.59$ &
  \multicolumn{1}{l|}{$1.68 - 5.76$} &
  \multicolumn{1}{l|}{$\leq 1.68$} &
  $\leq 0.73$ \\ \hline
\begin{tabular}[c]{@{}l@{}}Measured $R_{att}$ in \\ Charger 1 using \\ simulated vehicle load ($k\Omega$)\end{tabular} &
  \multicolumn{1}{l|}{\begin{tabular}[c]{@{}l@{}}B $\leftarrow$ B $\rightarrow$ C: $0.145$\\ A $\leftarrow$ B $\rightarrow$ C: $3.78$\end{tabular}} &
  B $\leftarrow$ C: $0.95$ &
  \multicolumn{1}{l|}{$1.9 - 5.4$} &
  \multicolumn{1}{l|}{-} &
  - \\ \hline
\begin{tabular}[c]{@{}l@{}}Measured $R_{att}$ in \\ Charger 2 using \\ simulated vehicle load ($k\Omega$)\end{tabular} &
  \multicolumn{1}{l|}{\begin{tabular}[c]{@{}l@{}}B $\leftarrow$ B $\rightarrow$ C: $0.392$\\ A $\leftarrow$ B $\rightarrow$ C: $2.92$\end{tabular}} &
  \begin{tabular}[c]{@{}l@{}}C $\leftarrow$ C $\rightarrow$ F: $0.52$\\ B $\leftarrow$ C $\rightarrow$ F: $1.1$\end{tabular} &
  \multicolumn{1}{l|}{$1.64 - 4.7$} &
  \multicolumn{1}{l|}{$\leq 1.5$} &
  $\leq 0.618$ \\ \hline
\begin{tabular}[c]{@{}l@{}}Measured $R_{att}$ in \\ Charger 1\\ on Tesla ($k\Omega$)\end{tabular} &
  \multicolumn{1}{l|}{$2.95$} &
  $0.72$ &
  \multicolumn{1}{l|}{$1.8 - 5.0$} &
  \multicolumn{1}{l|}{$\leq 0.93$} &
  $\leq 0.01$ \\ \hline
\begin{tabular}[c]{@{}l@{}}Measured $R_{att}$ in \\ Charger 2\\ on Tesla ($k\Omega$)\end{tabular} &
  \multicolumn{1}{l|}{$2.30$} &
  $0.50$ &
  \multicolumn{1}{l|}{$2 - 5.3$} &
  \multicolumn{1}{l|}{$\leq 0.5$} &
  $\leq 0.25$ \\ \hline
\begin{tabular}[c]{@{}l@{}}Measured $R_{att}$ in \\ Public Charger \\ on Tesla ($k\Omega$)\end{tabular} &
  \multicolumn{1}{l|}{$2.02$} &
  $0.83$ &
  \multicolumn{1}{l|}{$1.5 - 5.8$} &
  \multicolumn{1}{l|}{$\leq 0.85$} &
  $\leq 0.31$ \\ \hline
\end{tabular}%
}
\end{table*}

\subsection{Exploring State Switching Boundary}
We first try to identify the state switching boundary for Charger 1 and 2. We connect the CP Pin with a diode and a potentiometer in a serial mode with the simulated vehicle load, then link it to the GND Pin to form a close circuit. The diode is used for blocking the negative voltage and the potentiometer is used to adjust the resistance of the connected vehicle load. 
By adjusting the connected resistance, we record the state switching voltage in Table~\ref{tab:state_switch_boundary}. The results show that Charger 1 and Charger 2 have the same state transition voltages for the state switching from A to B, and B to C. However, Charger 1 does not have state F (omitted by the manufacturer), 
whereas the corresponding high pilot voltage of state transition for Charger 2 is $4.4V$. 
We estimate the operation range of $R_{att}$ of serial insertion attack using Eqs.~(\ref{equ1}), (\ref{equ2}), (\ref{equ3}) as shown 
in Section \ref{Sec:Attack_Design}, and that of parallel attachment attacks using  Eqs.~(\ref{equ4}), (\ref{equ5}), (\ref{equ6}). We summarize the estimated $R_{att}$ working range in Table~\ref{tab:serial_parallel_result}. 


\subsection{Performance of Serial Insertion Attack in the Lab}
The serial insertion attack circuit includes a potentiometer in parallel with a switch 
for easier resistance adjustment. 
We first launch the serial attack in a lab environment. Specifically, we design and implement a simulated vehicle load, \emph{i.e.}, a resistor connected to a diode, which is the same as the EV load shown in Figure \ref{fig:attack_circuit_design}. In the experiment, the switch is initially closed such that $R_{att}$ is not connected to the control pilot due to the short circuit. Once the attack starts, 
we will keep the switch open and start rotating the potentiometer from 0$\Omega$ to the value that can cause the state to switch. 

The experimental results are shown in Table~\ref{tab:serial_parallel_result}. 
$\leftarrow or \rightarrow$ stands for the direction of state switching on EVSE and EV sides, respectively. For example, $A \leftarrow B \rightarrow C$ in the serial insertion attack denotes that 
the state switches from B to A on the EVSE end, and B to C on the EV end. 
This is because, with the increasing $R_{att}$, the voltage difference between $V_{EVSE}$ and $V_{EV}$ also grows, \emph{i.e.}, 
$V_{EVSE}$ slowly approaches $V_{A/B}$, while $V_{EV}$  approaches $V_{B/C}$. Note that $V_{EVSE}$ and $V_{EV}$ may not reach the boundary at the same time. 
Moreover, $V_{EVSE}$ and $V_{EV}$ stay the same
during the parallel attachment attacks, and therefore in this case, they will read the same pilot signals regardless of the value of $R_{att}$.

We test our attack when the initial states are in state B and C. When we launch our attack on state B, we observe that $V_{EV}$ will first cross $7.8V$, which is the transition voltage from state B to state C, while the $V_{EVSE}$ is still in the range of state B. For Charger 1 and Charger 2, the transition resistances are mesured as $0.145k\Omega$ and $0.392k\Omega$, respectively. 
The estimated $R_{att}$ range is an upper bound that can guarantee the success of the attack. As the resistance increases, $V_{EVSE}$ exceeds $10.4V$, 
when the charger switches to state A. We note that both $3.78k\Omega$ and $2.92k\Omega$ for Charger 1 and 2 (from state B to C) are within the estimated resistance range. The attack on state C has similar performance.

\subsection{Performance of Serial Insertion Attack on Tesla}
In the real-world attack, we connect a Tesla Model 3 in the place of the simulated vehicle load and repeat the attacks towards the chargers. We test the attacks on Charger 1, Charger 2, and Public Charger. 
Here, we only record the resistance once the communication error is detected. 
Figure \ref{state_switch_CtoD} shows that the communication error is detected by Tesla, which immediately halts the charging process. When the Tesla is in error state, we further adjust $R_{att}$ to a smaller value trying to return to a normal state. 
However, we find that once the error is detected by the EV, returning back to the normal state is impossible, 
unless we re-plug in the charger's connector. This is likely a fail-safe feature of the vehicle. 

In the first experiment, we test the scenario of $A \leftarrow B \rightarrow C$, when we attempt to switch the EV to state C and EVSE to state A. Yet, due to the BMS function, the vehicle could not be switched to state C regardless of the parameter settings. Fortunately, by increasing the potentiometer, the attacker could switch the charging state of EVSE to state A. With a disparity of the state between EV and EVSE, a communication error occurs, resulting in a DoS. 


In the second experiment, we test $B \leftarrow C \rightarrow F$. Based on our measurement studies, when the high pilot voltage becomes low enough, the vehicle will switch to state F, \emph{i.e.}, error state. The increasing resistance of the potentiometer could eventually reduce the high pilot voltage to cause the state switching. Meanwhile, when the value of the resistance grows, EVSE will switch from state C to B, in which case a communication error will be observed. We find that both errors, \emph{i.e.}, ``low pilot voltage error" and ``communication error", could possibly occur, when the Tesla falls into state F. \emph{In summary, the serial insertion attack could cause DoS to the EV by causing ``low pilot voltage error" or ``communication error"}. 






\subsection{Performance of Parallel Attachment Attack in the Lab}
As shown in Figure~\ref{fig:parallel_design}, the parallel attachment attack circuit contains a potentiometer in serial with a switch and a diode. 
We first launch the parallel attachment attack in the lab environment with the simulated vehicle load. In the experiments, the switch is initially set to open, such that $R_{att}$ is not connected to the pilot line due to the open circuit. 
Once the attack starts, we will close the switch and start rotating the potentiometer from its maximum of $10k\Omega$ to the value that can cause the state to change.

We record the results in Table~\ref{tab:serial_parallel_result}. Similar to the serial insertion attack, we test our attack in states B and C. For Charger 1, when we launch the attack to switch the state from B to C, we observe the state transition and record the resistance $5.4k\Omega$ as the maximum $R_{att}$. Then, we keep adjusting the potentiometer until the state jumps to state F, and we record its value $1.9k\Omega$ as the minimum $R_{att}$. 
Similarly, we launch the attacks for $C\rightarrow F$ and $B\rightarrow F$. We find that the measured resistance range falls within or matches with the estimated range in all three attack scenarios. 

\subsection{Performance of Parallel Attachment Attack on Tesla}
We repeat the same attack process and record the resistance range on a Tesla. We find that the measured $R_{att}$ range is very similar to the estimated one in the attack $B \rightarrow C$. In the attacks $C \rightarrow F$ and $B \rightarrow F$, we successfully launch the attack on all the three EV chargers with an appropriate $R_{att}$. 
Charger 2 directly switches to state F, disconnecting itself from the EV. 
However, Charger 1 and Public Charger do not possess state F, yet, the state switching (\emph{i.e.}, from C/B to F) at the EV end still occurs. 
The experimental outcomes on Public Charger show nearly identical results with the two home chargers. 


Note that the manual adjustment of the potentiometer and the responding delay of the Tesla could cause a reduction of the boundary accuracy, which interprets the difference between measured values and the estimated ones.
\emph{In summary, the parallel attachment successfully achieves state switching on the Tesla vehicle.} 



\subsection{Performance of Automation Attack}
The automation attack does not require human involvement (manually turning the switch to start or terminate the attack). The state will be switched automatically from a certain state to another state. In the lab environment, we test it by adjusting the simulated vehicle load and check whether the charger will switch to the planned states. Using our attack circuit, we successfully switch the charger state from B to C and C to F on both Charger 1 and 2.

We test the automation attack on the Tesla when it connects with the Public Charger. Generally, it is challenging to fine-tune the attack parameters for supporting different chargers and vehicle loads. \emph{The results show that our attack can successfully switch the Tesla charging state from B to C, while it fails in switching its state from C to F. }

\begin{figure}[t]
\centering
\subfigure[Benign state B]{\includegraphics[width=0.24\textwidth]{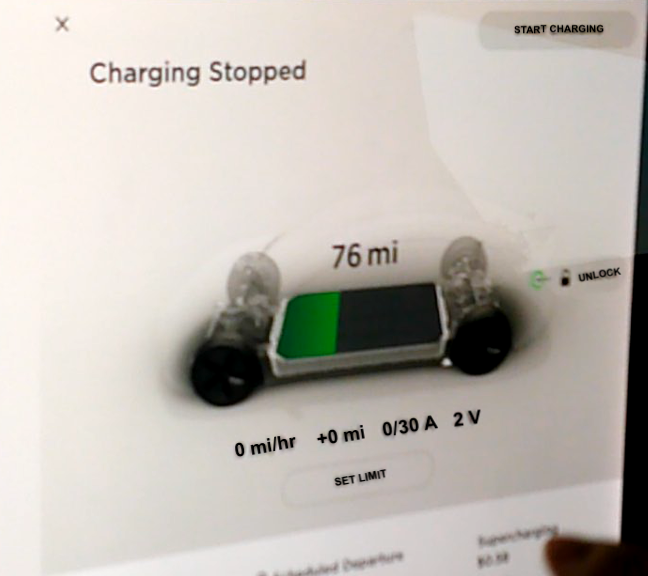}\label{benign_state_B}}
\subfigure[Benign state C]{\includegraphics[width=0.24\textwidth]{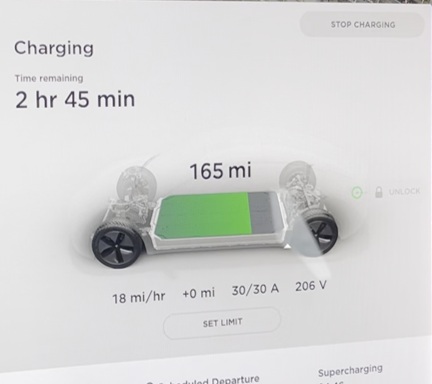}\label{benign_state_C}}
\caption{(a) and (b) shows the control panel of the Tesla Model 3 at state B and C without attack when it is connected to the Public Charger.}
\label{fig:without_attack_demo}
\end{figure}

\subsection{Attack Demonstration from the Tesla Control Panel}

\begin{figure}[htp]
\centering
\subfigure[Switch state B to C]{\includegraphics[width=0.44\textwidth]{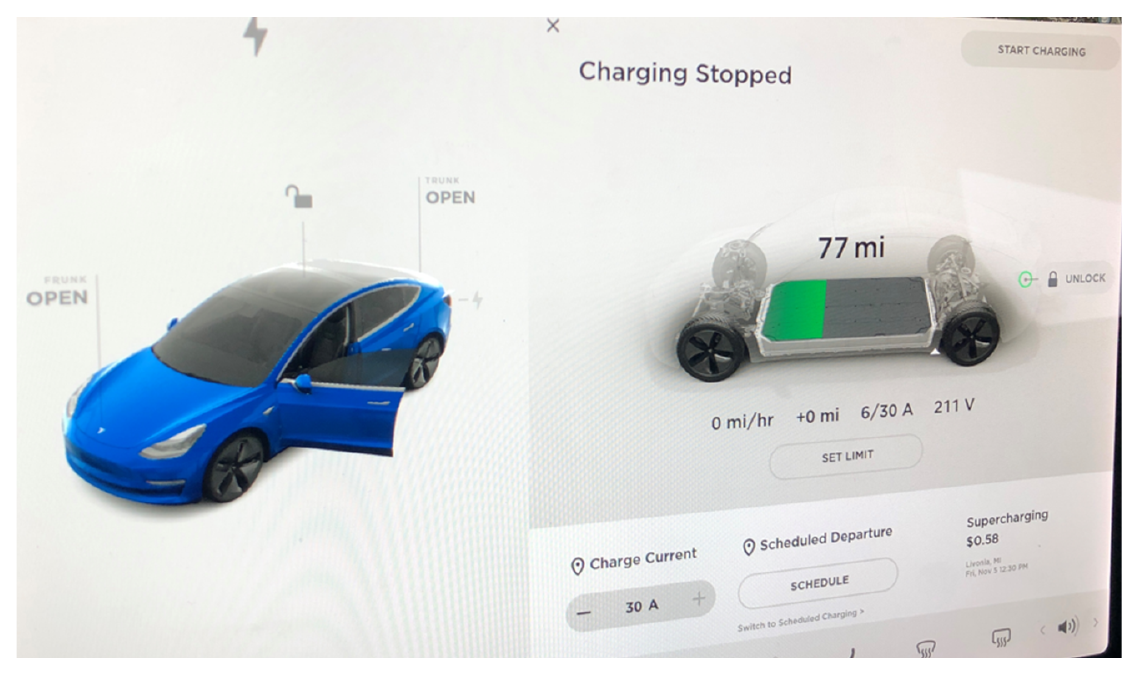}\label{state_switch_BtoC}}
\subfigure[Switch state B/C to F]{\includegraphics[width=0.44\textwidth]{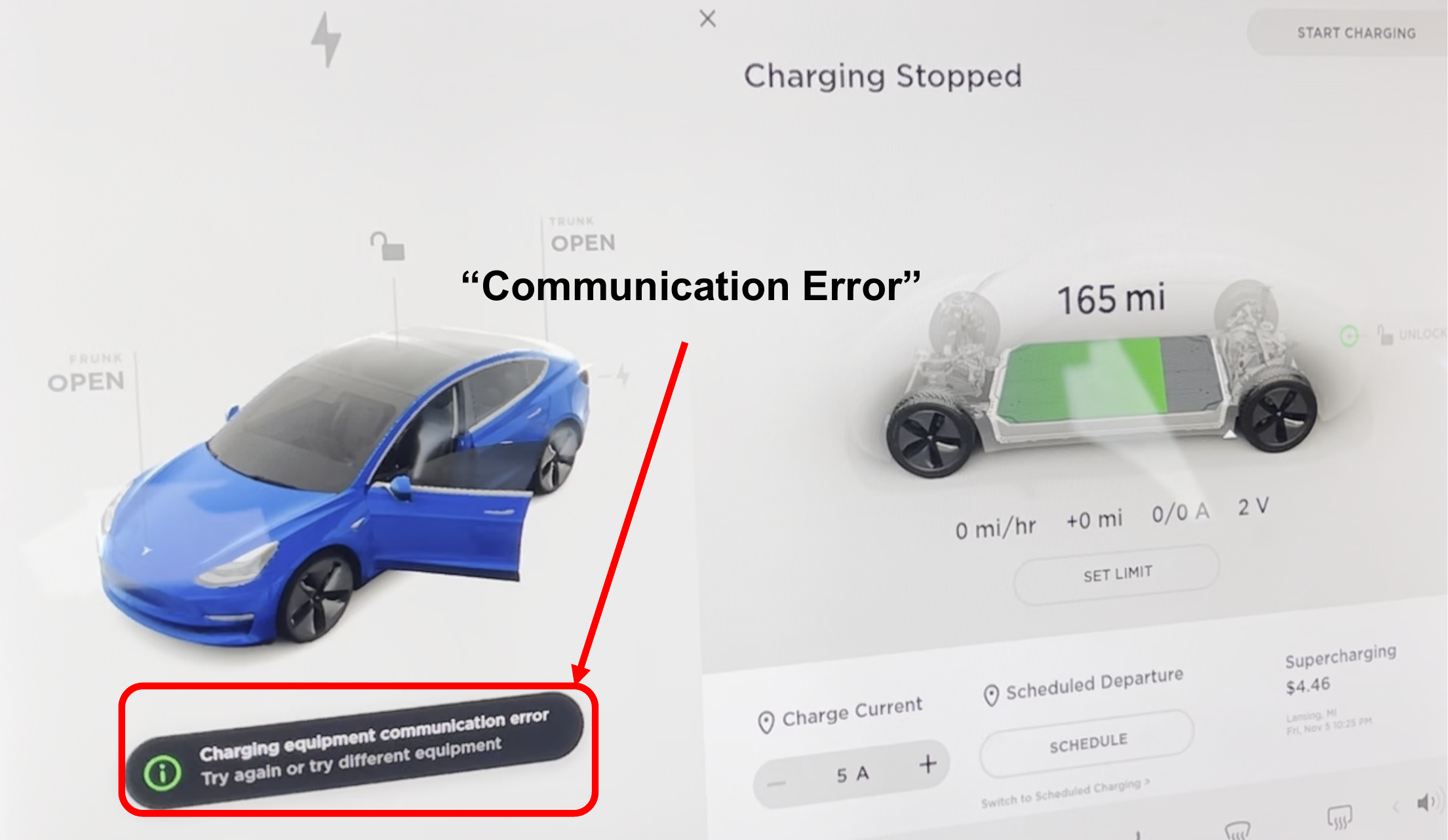}\label{state_switch_CtoD}}
\caption{(a) and (b) shows the state switching from B to C and B/C to F under cable tampering attacks as displayed on the control panel of Tesla Model 3 when connected to the Public Charger.}
\label{fig:with_attack_demo}
\end{figure}

We further demonstrate the attack results on the Tesla control panel. Figure \ref{benign_state_B} shows that the charging process is stopped manually by the human driver (\emph{i.e.}, clicking ``stop charging" button), and the vehicle is ready to charge (\emph{i.e.}, in state B) with the supply current and voltage readings as $0/30A$ and $2V$. 
However, when we launch our attack to switch the charging state from B to C,
as shown in Figure \ref{state_switch_BtoC}, the Public Charger continues to offer $6/30A$ and $211V$ for the charging activity regardless of the fact that the vehicle has stopped charging. 
On the Tesla control panel, the charging rate is displayed as +0 mile/hour,  whereas in the normal charging state C,  the panel (see Figure~\ref{benign_state_C}) shows the vehicle's charging rate is +18 miles/hour. 

\begin{figure}[htp]
\centering
\includegraphics[width=0.24\textwidth]{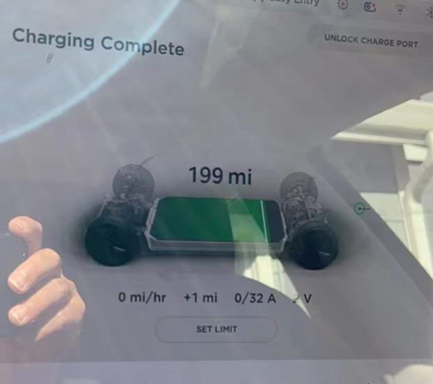}
\caption{Fully charged status of Tesla Model 3.}
\label{fig:fully_charged}
\end{figure}

Although the charging rate is displayed as +0 mile/hour, the current and voltage (\emph{i.e.}, $6/30A$ and $211V$) induced by our attack has to be consumed by the vehicle's components. However, without measuring the internals of the vehicle, it is impossible to know the exact consumers of these powers. 
Therefore, we suppose that there may be two possible outcomes of the attack. 
First, the battery is not being charged due to the existence of the BMS, and the power from the charger is consumed by other components on the vehicle such as heater unit, air conditioner unit, etc. Second, the battery level will eventually increase after a long period of time, e.g., 30 min. This indicates an over-charging situation when the BMS fails to react to the erroneous state switching. As shown in Figure~\ref{fig:fully_charged}, when the vehicle is fully charged, the voltage is $2V$ and the current is $0/32A$. However, if we launch our attack, the current and voltage may increase which lifts the battery level as the case in Figure~\ref{state_switch_BtoC}. If our attack causes the battery to be overcharged, one of the serious consequences could be battery damage, especially when the battery is already fully charged to 100\%. Moreover, the users would usually set a charge limit to optimize the battery life~\cite{optimize_battery, Every_Amp}. With our attack, the battery level may be charged over the set limit, causing a reduction of the battery capacity in a long term.

We consider that the first scenario is a more likely outcome for Tesla car, as the Tesla BMS could prevent the overcharging situation to occur. A more refined examination of Tesla BMS~\cite{Patrick_Kiley} would provide the support for an overcharging attack opportunity with the state switching attacks, and we leave such investigation for future work. 


\section{Exploring Charging Rate Attack}
The above state switching attacks demonstrate the attack capability in manipulating the charging states. The duty cycle of the PWM signal also plays an important role in EV charging, since it determines the charging rate. In this section, we explore the Charging Rate Attack, which aims to reduce the EV charging rate. Since the charging rate is proportional to the duty cycle of the PWM signal based on Table~\ref{duty_cycle_table}, reducing the charging rate requires control over the duty cycle of the PWM signal. 

To achieve the attack goal, we design two different attack circuits. One is to use a monostable TLC555 circuit to shrink the duty cycle. The other design involves a fake vehicle load. By using a comparator and 4 MOSFETs as switches, the duty cycle of the PWM signal on the vehicle side can be modified. One unique characteristic of our design is that these circuits do not need external power supplies, making them suitable to be integrated into a malicious extension cable in an unnoticeable manner. Both designs succeed in decreasing the charging rate with proper attack parameters in the lab environment.

\subsection{TLC555-based Attack Design}


\begin{figure}[htp]
\centering
\includegraphics[width=0.43\textwidth]{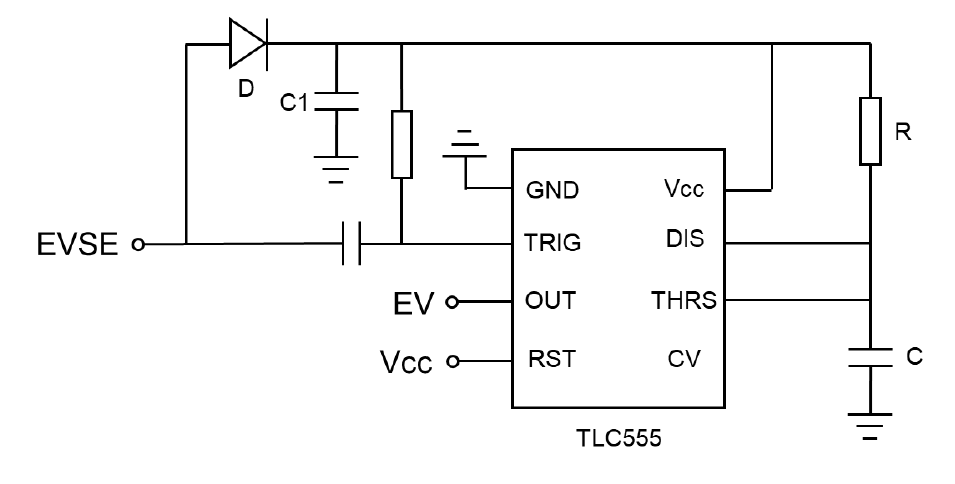}
\caption{The design of the TLC555-based duty cycle attack circuit.}
\label{fig:duty_design_1}
\end{figure}

Figure \ref{fig:duty_design_1} shows a high-level overview of the TLC555-based duty cycle attack design. We connect the attack circuit in serial with the vehicle, aiming to change the duty cycle of the PWM signal in state C. In this attack, we configure 
$V_{EVSE}=V_{EV}=6V$ such that both the charger and vehicle will be in state C. 
To avoid adding an extra power source, 
we convert the PWM signal to a constant $6V$ DC voltage using the capacitor $C1$ and diode $D$ to power up the TLC555 timer. The rest of the circuit is a monostable TLC555 circuit~\cite{555_Monostable} designed to reduce the duty cycle, e.g., from 25\% to 10\%. The circuit is triggered on a negative-going pulse, and the output of the circuit is a square pulse wave with ``high" and ``low" states. 
Once triggered, it will remain in 
a ``high" output state until a time period ($1.1 \cdot R \cdot C$) set up by the resistor $R$ and capacitor $C$ has elapsed. 
Thus, by changing the values of $R$ and $C$, the duty cycle can be adjusted accordingly.

\subsection{Fake Load-based Attack Design} 

The basic idea behind the fake load-design is to decrease the high pilot voltage to $-12V$, with MOSFETs acting as switches to toggle between the attack and non-attack periods. In the attack period, the real EV load is disconnected with the EVSE, whereas a fake vehicle load is connected instead.

\begin{figure}[htp]
\centering
\includegraphics[width=0.5\textwidth]{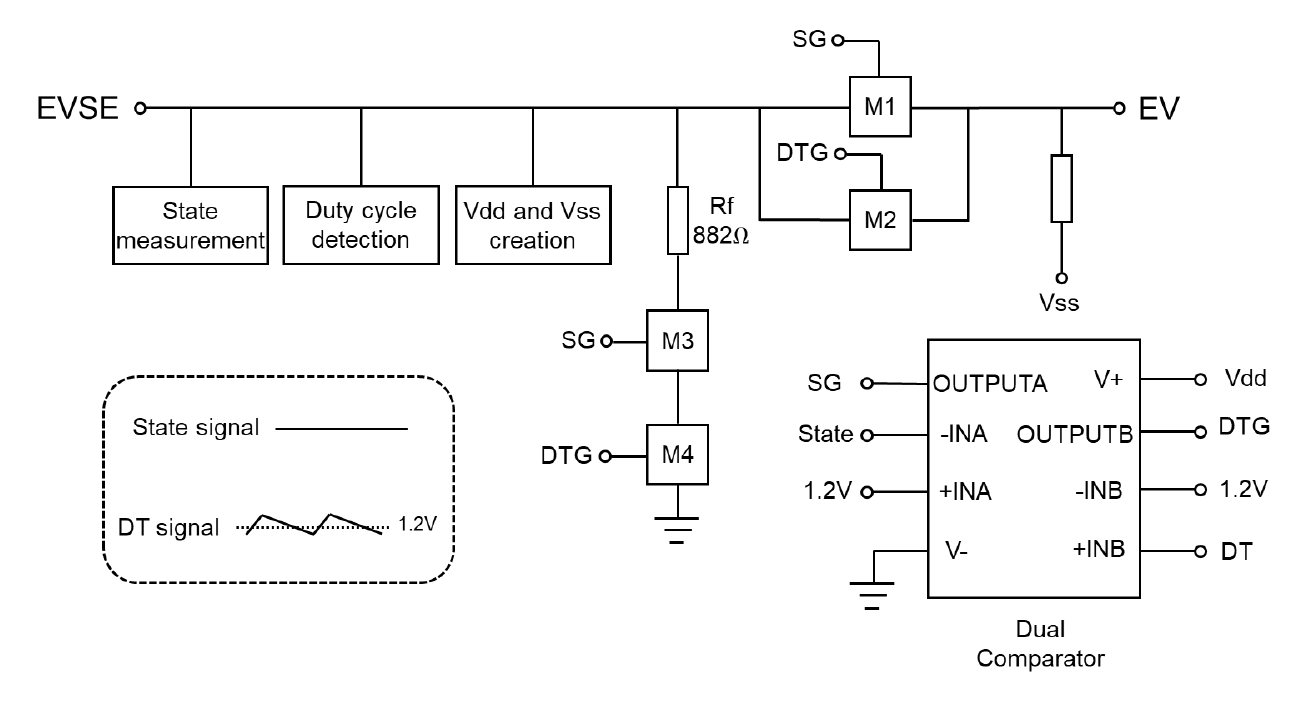}
\caption{The design of the fake load-based duty cycle attack circuit.}
\label{fig:duty_design_2}
\end{figure}

\begin{figure*}[t]
\centering
\subfigure[Duty cycle in benign state C]{\includegraphics[width=0.32\textwidth]{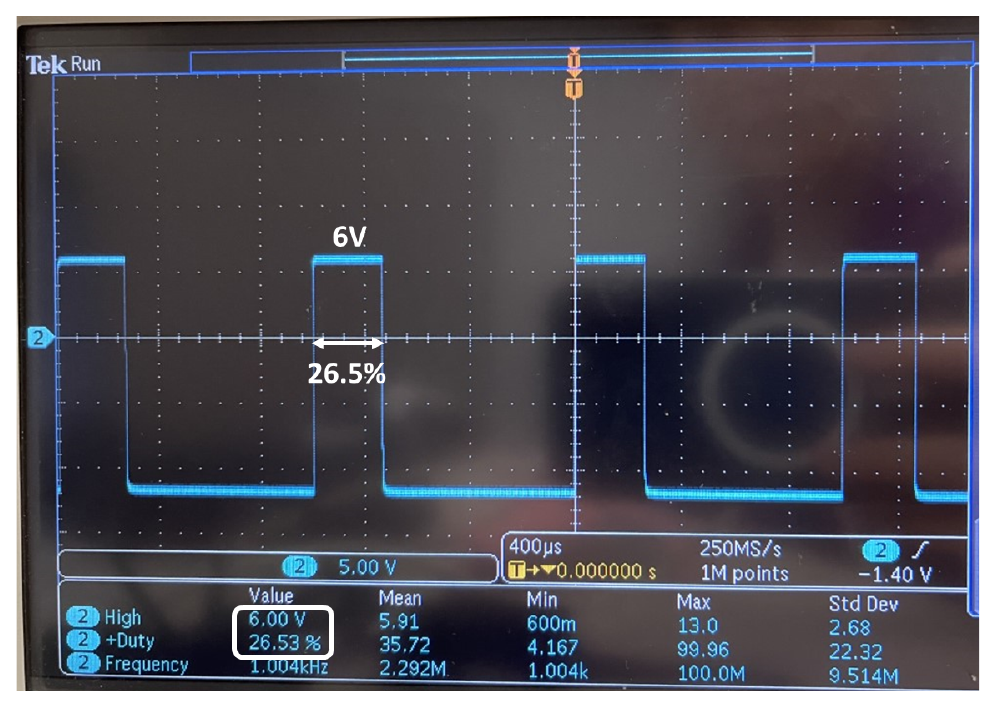}\label{DTresult_state_C}}
\subfigure[An example for TLC555-based duty cycle attack]{\includegraphics[width=0.32\textwidth]{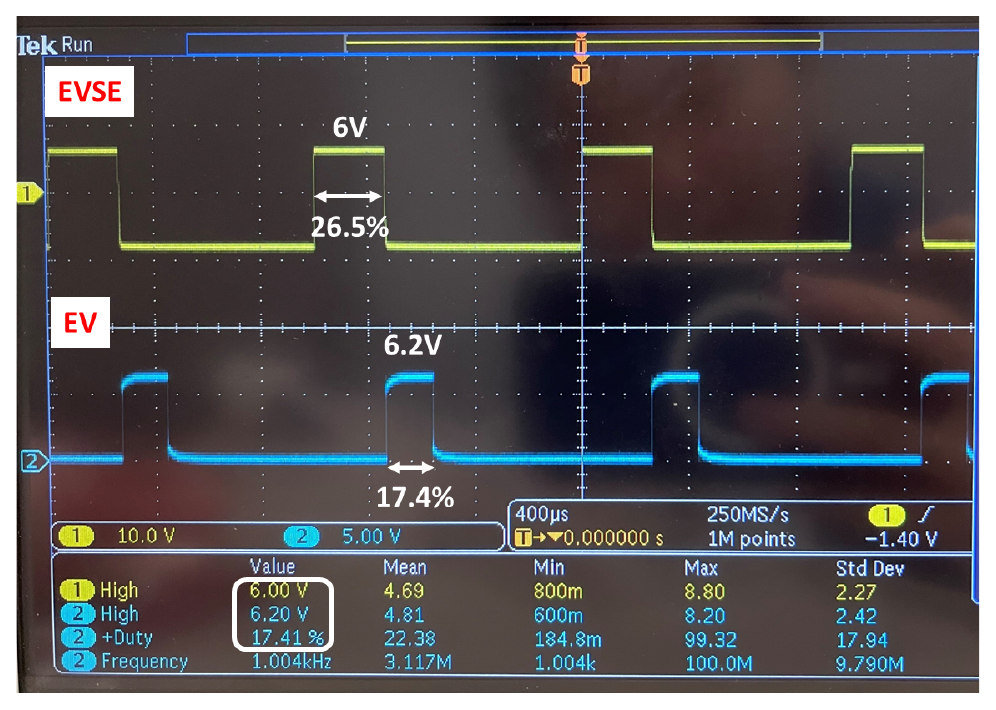}\label{DTresult_555}}
\subfigure[An example for fake load-based duty cycle attack]{\includegraphics[width=0.32\textwidth]{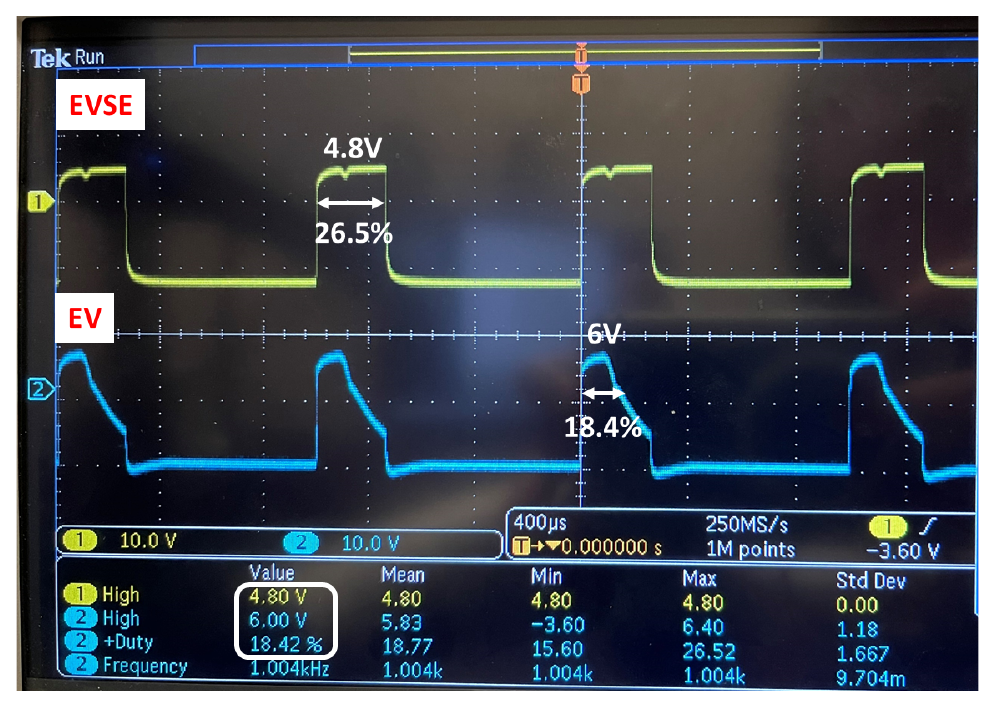}\label{DTresult_fakeload}}
\caption{(a) Duty cycle in state C without attack; (b)(c) Duty cycle with attacks using two different attack designs.}
\label{fig:duty_cycle_result}
\end{figure*}

As shown in Figure \ref{fig:duty_design_2},  fake load-based design contains four blocks, providing state measurement, duty cycle detection, $V_{dd}$ and $V_{ss}$ creation, and signal comparison \& duty cycle reduction. The first block is designed to detect the high voltage of the PWM signal from the charger using a RC circuit with a large time constant. 
It allows us to derive the ``State" signal information from the high pilot voltage. The second block detects the duty cycle of the PWM signal on the charger side using a RC circuit with a small time constant. The capacitor is continuously charged and discharged to create a triangle wave (\emph{i.e.}, 
Duty cycle control signal, DT) which can be used to detect the duty cycle of the PWM signal based on how fast it ramps up or down. The third block creates a $V_{dd}$ corresponding to the state peak value, and a $V_{ss}$ with the value of $-12V$ using a diode and a capacitor. The forth block consists of a dual comparator and 4 MOSFETs. The dual comparator compares the State and DT with the $1.2V$ DC reference voltage to output the State Gate (SG) and DT Gate (DTG), which are the gate values used to decide whether to turn on/off the 4 MOSFETs. As a result, the fake load $R_f$ will be connected or disconnected based on the charging state. 

By adjusting the parameters, we first aim to disable the attack in state A and B, while enabling the attack in state C. MOSFET M1 and M2 will be turned on when we apply $0V$. MOSFET M3 and M4 will be turned on when we apply $V_{dd}$. Therefore, we adjust State signal to be higher than $1.2V$ in state A and B, and lower than $1.2V$ in state C. 
Since M1 is turned on and M3 is turned off in state A and B, the EVSE is always connected with the real EV load. In state C, however, 
M1 is turned off and M3 is turned on.  
Therefore, M2 and M4 will further determine the connection between EV and EVSE.

The DT signal reflects the duty cycle of the PWM signal. Since it controls the value of DTG, its voltage level decides the status of M2 and M4. By adjusting the parameters, part of the DT signal is above $1.2V$ while the rest  is below $1.2V$. When DT is below $1.2V$, DTG is $0V$ which turns on M2 and turns off M4. This allows the charger to still connect with the EV. On the contrary, when DT is above $1.2V$, DTG is $V_{dd}$ which turns off M2 and turns on M4. Then, the attack is activated, causing EV to detach from the EVSE and connect to $V_{ss}$. As a result, it changes the PWM signal on the EV side from $6V$ to $V_{ss}$, resulting in the decrease of the duty cycle. On the other hand, when EVSE is connected to the fake load $R_f$, it stays in state C without being interrupted. 
The DT signal can be tweaked by changing the values in the second block, allowing us to change the duty cycle to different values.

\subsection{Performance of TLC555-based Attack}
We perform the two duty cycle attacks in both a lab environment and a real-world physical environment 
with a Tesla car. 
We first evaluate the TLC555-based duty cycle attack circuit in a lab environment using the simulated vehicle load. Figure \ref{DTresult_state_C} shows the duty cycle in state C without attack. It can be seen that the high voltage level of the signal is $6V$ and its duty cycle is 26.5\% which is corresponding to the charging rate of $16A$. Then, we launch our attack towards the Charger 2. The duty cycle can be changed to different values by adjusting the $R$ and $C$ value in Figure~\ref{fig:duty_design_1}. 

Figure \ref{DTresult_555} shows an example of the duty cycle attack. The signal in yellow is the PWM signal read by the EVSE. Its high voltage is $6V$ and the duty cycle is 26.5\%. The blue 
signal on oscilloscope is the output PWM signal at the EV side. The high voltage level is $6.2V$ and the duty cycle is 17.41\% which corresponds to the charging rate of $10A$. We can see that the charging rate is reduced by $6A$. It demonstrates the success of our attack in the lab environment. 

However, the attack fails on 
the Tesla car. The reason is that the duty cycle attack circuit starts functioning once it is connected to the charger, which means that state A is absent on the EV side, which leads to a failure/error state. 
To address the problem in real-world scenario, we can implement remotely-controlled switches in the attack circuit design to control the starting point of the attack so that the attack will not be activated in state A and B. 



\subsection{Performance of Fake Load-based Attack}
We test the fake load-based duty cycle attack circuit in the lab. The duty cycle can be set to different values by adjusting the resistance in the second block, as it controls the charging rate of the RC circuit. Figure \ref{DTresult_fakeload} displays an example of duty cycle signals on an oscilloscope. 
The signal in yellow is the PWM signal at the charger side. The corresponding high voltage level is $4.8V$ and the duty cycle is 26.5\%. The blue signal is the output PWM at the vehicle side, whose high voltage level is $6V$ and the duty cycle is 18.42\% which corresponds to the charging rate of $11A$. 
We can see that the charging rate is reduced by $5A$, which demonstrates the success of our attack. Since we are using the analog circuit, the voltage cannot drop from $6V$ to $V_{ss}$ immediately, which explains the slope of the blue signal. 

We further evaluate our attack on the Tesla. We first charge the vehicle in state C and initiate the attack. We observe that once our attack is activated, the Tesla starts displaying the communication error. The reason might be attributed to the imperfect slope of the PWM signal on the vehicle side. One solution is to use a microcontroller for generating fake duty cycles as has been shown in a recent research~\cite{SwRI_attack}. 
However, turning on the microcontroller necessitates the use of an external power source or a more complex power harvesting solution, which may further complicate the attack implementation and maintenance.



\section{Discussion}

In this section, we will discuss the impact of vehicle's BMS, as well as the practicality, generality, robustness, and stealthiness of the proposed attacks. We also discuss the limitations and the potential defense mechanisms for countering \attack attack.

\subsection{Battery Management System}
For a production-grade electronic vehicle, a BMS is usually configured for rechargeable batteries to monitor the state of the battery packs and protect them from high-risk operations. 
BMS is also designed to protect the battery pack from malicious attacks. However, previous studies have shown that potential vulnerabilities of BMS can lead to various cyber attacks (e.g., faulty components)~\cite{sripad2019vulnerabilities}. Cyber attacks could cause overdischarge or overcharge, which may damage the battery packs.
In severe cases, it may cause internal shorts and greatly shorten the lifetime of the battery pack~\cite{sripad2019vulnerabilities}. 
BMS is also vulnerable to IEMI attacks~\cite{dayanikli2020electromagnetic}. Experiments have shown that attacker can takeover the control of BMS by intercepting and manipulating the feedback signal of BMS and cause damage to the EV~\cite{dayanikli2020electromagnetic}.
In this work, we mainly focus on the functionality of the EV chargers. By utilizing the existing attacks towards BMS, we might be able to bypass the BMS and cause permanent damage to the EV battery. For example, switching the charging state from B to C might overcharge the battery if the BMS is not configured properly. 

\subsection{Practicality} 
In our attack, 
the attacker may connect the malicious extension cable to the public charging station secretly when the charger is not in use. Furthermore, the attacker can disguise the malicious extension cable's connector as a benign connector by hanging it on a public charging station.
To attack the home chargers, the attacker can secretly add the attack circuit into the extension cable. 
For instance, connecting the parallel attachment attack circuit to the extension cable does not affect the normal charging process when the attack switch is open.


\subsection{Generality}
We mainly test \attack on the AC charging systems, but we believe that \attack \textit{is also applicable to the DC charging}. 
According to the ISO 15118 standard, J1772 DC charging also employs the same PWM pilot control. This suggests that our proposed attack might be equally effective for DC charging. One caveat is that the attacker's  automation circuit may not work for switching from state B to C. As the attack circuit directly switches state B to C without any time delay, it may not be able to bypass the required digital communication process in state B of DC charging. 

\subsection{Robustness}
In \attack, we mathematically calculate the attack resistance range in Section~\ref{Sec:Attack_Design}. The results show that for different EV chargers, the estimated attack resistance range is very similar to each other. Meanwhile, we prove that the estimated attack resistance range is close to the measured one in both the simulated vehicle load case and the real vehicle case. The attacker can select the resistance in the estimated working range, and directly apply the attack to different chargers to achieve a plausible success rate. 

\subsection{Stealthiness}
In our attack, the attacker can hide the attack in the extension cable's adapter, thanks to the tiny size of the attack circuit. For example, the serial insertion attack contains only one resistor, which is as small as a quarter coin. The automation circuit in our design has the most number of components, however, its size is smaller than a typical bank card. Furthermore, because we assemble the circuit by hand, the component positioning is not optimized; nevertheless, smaller components could be used if the PCB was assembled by a pick-and-place machine, resulting in an even more miniature attack circuit.

\subsection{Limitation}
The first limitation lies in that the attacker needs to be involved to manually initiate and terminate the attack in serial insertion and parallel attachment attacks.
The second limitation is that our charging rate attacks do not change the charging current on a Tesla vehicle likely due to the slower response time of the analog circuit.
However, it is possible to change the charging duty cycles by using a microcontroller, although the involvement of a microcontroller would require an additional power source. 
The third limitation is that accessing a home charger may be challenging in real life. EV users usually charge their vehicles inside the yards or garages of residential houses. It may be hard to get physical access to the home charger if the charging environment is not publicly accessible. 

\subsection{Defense}
There are several defense mechanisms to defend against our attack. 
The users could avoid using the extension cables, or should meticulously inspect them before use to ensure they have not been opened, altered, or modified. If the attack circuit is implemented in other weak points of the chargers, it is important for users to check the chargers comprehensively before charging. For the EV manufacturer, we recommend a better pre-check procedure to be implemented, e.g., the EVs should always check the state consistency between the EV and EVSE. 
The EVs should also have tighter ranges for states B, C, and F. Once EVs are locked in a state, they should constantly monitor the pilot voltage to make sure that the state or voltage is not drifting. They should also have tighter tolerance for $-12V$ measurement. These two measures would significantly increase the difficulty of a circuit design that allows to harvest power from the communication signals, and thus increase the complexity of a successful attack.

\section{Related Work}

In this section, we present the most relevant work regarding the security of EV charging and power grids.


\subsection{Security of EV Charging} The security research of EV charging is a newly-emerging area. 
Dayanikli et al.~\cite{dayanikli2020electromagnetic} exploit the nonlinearity of amplifiers and analog-to-digital converters (ADCs), which are commonly used in power converters for sensing and feedback control, to attack DC charging. The adversary is able to manipulate the voltage and current outputs which is necessary to ensure the proper operation of the converter by injecting IEMI. Baker et al.~\cite{baker2019losing} propose a passive data eavesdropping method towards the DC charging via digital communication, \emph{i.e.,} PLC. The attacker listens to the unintended electromagnetic radiation of the EV charging communication and develops a PLC eavesdropping tool to eavesdrop on private data. 
Different from the existing studies, we demonstrate the active attack toward PWM pilot control used by both AC and DC charging.  
Moreover, our attack achieves the precise manipulation of the charging state.

\subsection{Security of Power Grids} There are also some recent studies  aiming to attack power grids. Antonakakis et al.~\cite{antonakakis2017understanding} compromise over 600,000 IoT devices to launch the DDoS attack to paralyze the network infrastructures. Going beyond traditional computer network infrastructures, Soltan et al.~\cite{soltan2018blackiot} show that attacking hundreds of thousands of high-energy IoT devices (such as water heaters and air conditioners) can disrupt the power grid in a variety of ways, e.g., line failures and operating costs. By controlling EV charging, Sayed et al.~\cite{sayed2022electric} show that EV loads have a high reactive power demand, which can have a greater effect on the grid than residential loads when attacking the power grid compared to earlier work. To defend against power grid attacks, Huang et al.~\cite{huang2019not} demonstrate that more protections on the transmission grid operation can enable the system to withstand a wide range of attacks and prevent a system blackout.

\section{Conclusion}
This paper presents \attack, which is comprised of state switching attack and charging rate attack. These attacks are engineered to manipulate the PWM pilot control across diverse EV charging systems, thereby enabling state switching and altering the charging rates. 
The proposed attacks actively manipulate the charging states, leading to dangerous outcomes ranging from charging fraud to battery damage. We perform the attack evaluation on a public AC charging station and two home chargers, using a simulated vehicle load as well as a Tesla Model 3. 
Our experimental demonstration shows that both the state switching attack and the charging rate attack can successfully manipulate the charging status, thereby causing disruption in EV charging.

\bibliographystyle{IEEEtran}
\bibliography{reference}

\begin{thebibliography}{10}
\providecommand{\url}[1]{#1}
\csname url@samestyle\endcsname
\providecommand{\newblock}{\relax}
\providecommand{\bibinfo}[2]{#2}
\providecommand{\BIBentrySTDinterwordspacing}{\spaceskip=0pt\relax}
\providecommand{\BIBentryALTinterwordstretchfactor}{4}
\providecommand{\BIBentryALTinterwordspacing}{\spaceskip=\fontdimen2\font plus
\BIBentryALTinterwordstretchfactor\fontdimen3\font minus
  \fontdimen4\font\relax}
\providecommand{\BIBforeignlanguage}[2]{{%
\expandafter\ifx\csname l@#1\endcsname\relax
\typeout{** WARNING: IEEEtran.bst: No hyphenation pattern has been}%
\typeout{** loaded for the language `#1'. Using the pattern for}%
\typeout{** the default language instead.}%
\else
\language=\csname l@#1\endcsname
\fi
#2}}
\providecommand{\BIBdecl}{\relax}
\BIBdecl

\bibitem{tie2013review}
S.~F. Tie and C.~W. Tan, ``A review of energy sources and energy management
  system in electric vehicles,'' \emph{Renewable and sustainable energy
  reviews}, vol.~20, pp. 82--102, 2013.

\bibitem{larminie2012electric}
J.~Larminie and J.~Lowry, \emph{Electric vehicle technology explained}.\hskip
  1em plus 0.5em minus 0.4em\relax John Wiley \& Sons, 2012.

\bibitem{editorial2020electricbus}
\BIBentryALTinterwordspacing
Editorial, ``Electric bus, main fleets and projects around the world,'' May
  2020. [Online]. Available:
  \url{https://www.sustainable-bus.com/electric-bus/electric-bus-public-transport-main-fleets-projects-around-world/}
\BIBentrySTDinterwordspacing

\bibitem{zoe2020california}
\BIBentryALTinterwordspacing
Z.~Woodcraft, ``California passes nation’s first electric trucks standard,''
  June 2020. [Online]. Available:
  \url{https://earthjustice.org/news/press/2020/california-passes-nations-first-electric-trucks-standard}
\BIBentrySTDinterwordspacing

\bibitem{electric2020construction}
\BIBentryALTinterwordspacing
E.~V. Research, ``Construction industry's first fully electric backhoe
  loader,'' March 2020. [Online]. Available:
  \url{https://www.electricvehiclesresearch.com/articles/20129/construction-industrys-first-fully-electric-backhoe-loader}
\BIBentrySTDinterwordspacing

\bibitem{meticulous2021electric}
\BIBentryALTinterwordspacing
Meticulous, ``Electric vehicle (ev) market worth \$2,495.4 billion by 2027,
  growing at a cagr of 33.6\% from 2020- exclusive report by meticulous
  research,'' May 2021. [Online]. Available: \url{https://tinyurl.com/yc64es63}
\BIBentrySTDinterwordspacing

\bibitem{wallbox}
\BIBentryALTinterwordspacing
{Wallbox}, ``Ev charging current: What's the difference between ac and dc?''
  2021, accessed: 2021-11-20. [Online]. Available:
  \url{https://wallbox.com/en_us/faqs-difference-ac-dc}
\BIBentrySTDinterwordspacing

\bibitem{On_off_board}
\BIBentryALTinterwordspacing
{Texas Instruments}, ``Taking charge of electric vehicles – both in the
  vehicle and on the grid,'' 2020, accessed: 2021-11-20. [Online]. Available:
  \url{https://bit.ly/3nuQHKa}
\BIBentrySTDinterwordspacing

\bibitem{yan2016can}
C.~Yan, W.~Xu, and J.~Liu, ``Can you trust autonomous vehicles: Contactless
  attacks against sensors of self-driving vehicle,'' \emph{Def Con}, vol.~24,
  no.~8, p. 109, 2016.

\bibitem{nassi2020phantom}
B.~Nassi, Y.~Mirsky, D.~Nassi, R.~Ben-Netanel, O.~Drokin, and Y.~Elovici,
  ``Phantom of the adas: Securing advanced driver-assistance systems from
  split-second phantom attacks,'' in \emph{Proceedings of the 2020 ACM SIGSAC
  conference on computer and communications security}, 2020, pp. 293--308.

\bibitem{jin2022pla}
Z.~Jin, J.~Xiaoyu, Y.~Cheng, B.~Yang, C.~Yan, and W.~Xu, ``Pla-lidar: Physical
  laser attacks against lidar-based 3d object detection in autonomous
  vehicle,'' in \emph{2023 IEEE Symposium on Security and Privacy (SP)}.\hskip
  1em plus 0.5em minus 0.4em\relax IEEE Computer Society, 2022, pp. 710--727.

\bibitem{zhou2022doublestar}
C.~Zhou, Q.~Yan, Y.~Shi, and L.~Sun, ``$\{$DoubleStar$\}$:$\{$Long-Range$\}$
  attack towards depth estimation based obstacle avoidance in autonomous
  systems,'' in \emph{31st USENIX Security Symposium (USENIX Security 22)},
  2022, pp. 1885--1902.

\bibitem{xieaccess}
X.~Xie, K.~Jiang, R.~Dai, J.~Lu, L.~Wang, Q.~Li, and J.~Yu, ``Access your tesla
  without your awareness: Compromising keyless entry system of model 3.''

\bibitem{dayanikli2020electromagnetic}
G.~Y. Dayanikli, R.~R. Hatch, R.~M. Gerdes, H.~Wang, and R.~Zane,
  ``Electromagnetic sensor and actuator attacks on power converters for
  electric vehicles,'' in \emph{2020 IEEE Security and Privacy Workshops
  (SPW)}.\hskip 1em plus 0.5em minus 0.4em\relax IEEE, 2020, pp. 98--103.

\bibitem{selvaraj2018intentional}
J.~Selvaraj, ``Intentional electromagnetic interference attack on sensors and
  actuators,'' Ph.D. dissertation, Iowa State University, 2018.

\bibitem{tu2019trick}
Y.~Tu, S.~Rampazzi, B.~Hao, A.~Rodriguez, K.~Fu, and X.~Hei, ``Trick or heat?
  manipulating critical temperature-based control systems using rectification
  attacks,'' in \emph{Proceedings of the 2019 ACM SIGSAC Conference on Computer
  and Communications Security}, 2019, pp. 2301--2315.

\bibitem{baker2019losing}
R.~Baker and I.~Martinovic, ``Losing the car keys: Wireless phy-layer
  insecurity in {EV} charging,'' in \emph{28th {USENIX} Security Symposium
  ({USENIX} Security 19)}, 2019, pp. 407--424.

\bibitem{chargerratio}
\BIBentryALTinterwordspacing
C.~EVs, ``The charging port-to-ev ratio: residential versus commercial,''
  Accessed on July 11, 2021. [Online]. Available:
  \url{https://chargedevs.com/newswire/the-charging-port-to-ev-ratio-residential-versus-commercial/}
\BIBentrySTDinterwordspacing

\bibitem{TexasInstruments}
\BIBentryALTinterwordspacing
T.~Instruments, ``Design guide: Tida-010071 sae j1772-compliant electric
  vehicle service equipment reference design for level 1 and 2 ev charger,''
  2021, accessed: 2021-07-19. [Online]. Available: \url{https://bit.ly/3exgahE}
\BIBentrySTDinterwordspacing

\bibitem{extension_cable}
\BIBentryALTinterwordspacing
{Lectron Store}, ``J1772 extension cable,'' 2021, accessed: 2021-11-20.
  [Online]. Available: \url{https://amzn.to/3pj9doD}
\BIBentrySTDinterwordspacing

\bibitem{hall2017emerging}
D.~Hall and N.~Lutsey, ``Emerging best practices for electric vehicle charging
  infrastructure,'' \emph{Washington, DC: The International Council on Clean
  Transportation (ICCT)}, 2017.

\bibitem{SAEJ1772ChargingAdapter}
\BIBentryALTinterwordspacing
{Tesla}, ``Sae j1772 charging adapter,'' 2021, accessed: 2021-11-20. [Online].
  Available: \url{https://bit.ly/3exgahE}
\BIBentrySTDinterwordspacing

\bibitem{J1772_wiki}
\BIBentryALTinterwordspacing
Wikipedia, ``Sae j1772,'' 2021, accessed: 2021-07-22. [Online]. Available:
  \url{https://en.wikipedia.org/wiki/SAE_J1772}
\BIBentrySTDinterwordspacing

\bibitem{CCS_design}
\BIBentryALTinterwordspacing
{Initiative Charging Interface}, ``Design guide for combined charging system,''
  2015, accessed: 2021-11-20. [Online]. Available:
  \url{http://tesla.o.auroraobjects.eu/Design_Guide_Combined_Charging_System_V3_1_1.pdf}
\BIBentrySTDinterwordspacing

\bibitem{J1772_201710}
\BIBentryALTinterwordspacing
{SAE}, ``Sae electric vehicle and plug in hybrid electric vehicle conductive
  charge coupler j1772-201710,'' 2017, accessed: 2021-11-20. [Online].
  Available: \url{https://www.sae.org/standards/content/j1772_201710/}
\BIBentrySTDinterwordspacing

\bibitem{level1_charger}
\BIBentryALTinterwordspacing
{LEFANEV Store}, ``Lefanev ev charger,'' 2021, accessed: 2021-11-20. [Online].
  Available: \url{https://amzn.to/3JVB1ZW}
\BIBentrySTDinterwordspacing

\bibitem{level2_charger}
\BIBentryALTinterwordspacing
{MEGEAR Store}, ``Megear level 1-2 ev charger,'' 2021, accessed: 2021-11-20.
  [Online]. Available: \url{https://amzn.to/3qaWrKG}
\BIBentrySTDinterwordspacing

\bibitem{chargepoint}
\BIBentryALTinterwordspacing
{chargepoint}, ``chargepoint,'' 2021, accessed: 2021-11-20. [Online].
  Available: \url{https://www.chargepoint.com/}
\BIBentrySTDinterwordspacing

\bibitem{optimize_battery}
\BIBentryALTinterwordspacing
{Fred Lambert}, ``Tesla battery expert recommends daily charging limit to
  optimize durability,'' 2017, accessed: 2021-11-20. [Online]. Available:
  \url{https://electrek.co/2017/09/01/tesla-battery-expert-recommends-daily-battery-pack-charging/}
\BIBentrySTDinterwordspacing

\bibitem{Every_Amp}
\BIBentryALTinterwordspacing
{Every Amp}, ``Electric vehicle battery life best practices,'' 2021, accessed:
  2021-11-20. [Online]. Available:
  \url{https://everyamp.com/electric-vehicle-battery-life-best-practices/}
\BIBentrySTDinterwordspacing

\bibitem{Patrick_Kiley}
\BIBentryALTinterwordspacing
{Patrick Kiley}, ``Reverse engineering the tesla battery management system to
  increase power available,'' 2021, accessed: 2021-11-20. [Online]. Available:
  \url{https://tinyurl.com/3c7pu2d2}
\BIBentrySTDinterwordspacing

\bibitem{555_Monostable}
\BIBentryALTinterwordspacing
{Electronicsclub}, ``555 monostable,'' 2021, accessed: 2021-11-20. [Online].
  Available: \url{https://electronicsclub.info/555monostable.htm}
\BIBentrySTDinterwordspacing

\bibitem{SwRI_attack}
\BIBentryALTinterwordspacing
{SwRI}, ``Swri hacks electric vehicle charging to demonstrate cybersecurity
  vulnerabilities,'' 2020, accessed: 2021-11-20. [Online]. Available:
  \url{https://www.swri.org/press-release/electric-vehicle-charging-cybersecurity-vulnerabilities}
\BIBentrySTDinterwordspacing

\bibitem{sripad2019vulnerabilities}
S.~Sripad, S.~Kulandaivel, V.~Pande, V.~Sekar, and V.~Viswanathan,
  ``Vulnerabilities of electric vehicle battery packs to cyberattacks,'' 2019.

\bibitem{antonakakis2017understanding}
M.~Antonakakis, T.~April, M.~Bailey, M.~Bernhard, E.~Bursztein, J.~Cochran,
  Z.~Durumeric, J.~A. Halderman, L.~Invernizzi, M.~Kallitsis \emph{et~al.},
  ``Understanding the mirai botnet,'' in \emph{26th $\{$USENIX$\}$ security
  symposium ($\{$USENIX$\}$ Security 17)}, 2017, pp. 1093--1110.

\bibitem{soltan2018blackiot}
S.~Soltan, P.~Mittal, and H.~V. Poor, ``Blackiot: Iot botnet of high wattage
  devices can disrupt the power grid,'' in \emph{27th $\{$USENIX$\}$ Security
  Symposium ($\{$USENIX$\}$ Security 18)}, 2018, pp. 15--32.

\bibitem{sayed2022electric}
M.~A. Sayed, R.~Atallah, C.~Assi, and M.~Debbabi, ``Electric vehicle attack
  impact on power grid operation,'' \emph{International Journal of Electrical
  Power \& Energy Systems}, vol. 137, p. 107784, 2022.

\bibitem{huang2019not}
B.~Huang, A.~A. Cardenas, and R.~Baldick, ``Not everything is dark and gloomy:
  Power grid protections against iot demand attacks,'' in \emph{usenix security
  symposium}, 2019, pp. 1115--1132.

\end{thebibliography}



\ifCLASSOPTIONcaptionsoff
  \newpage
\fi

\end{document}